\documentclass{JHEP3}

\usepackage[T1]{fontenc}
\usepackage{amssymb,amsmath}
\usepackage{lmodern}

\DeclareMathOperator{\diam}{\diamond}
\DeclareMathOperator{\R}{Re}
\DeclareMathOperator{\I}{Im}
\newcommand{\ee}{\mathrm{e}}
\newcommand{\de}{\mathrm{d}}
\newcommand{\BH}{\text{bh}}
\newcommand{\Sa}{S}
\newcommand{\betad}{\alpha}

\newcommand{\Dbar}[2]{\bar{\mathcal{D}}_{\bar{#1}}\bar{#2}}
\newcommand{\Iprod}[2]{\langle {#1}, {#2} \rangle}
\newcommand{\db}{\boldsymbol{\de}}
\newcommand{\Omt}{\hat{\Omega}}
\newcommand{\Ftilde}{\tilde{\mathcal{F}}}
\newcommand{\Gtilde}{\tilde{\mathcal{G}}}
\newcommand{\Gamt}{\tilde{\Gamma}}
\newcommand{\boldeta}{\boldsymbol{\eta}}
\newcommand{\St}{S}
\newcommand{\Sb}{T}
\newcommand{\Ww}{\mathcal{W}}
\def\e{\epsilon}
\def\sF{{\boldsymbol{\cal F}}}
\def\cH{{\mathcal H}}
\def\cJ{{\mathcal J}}
\def\cR{{\mathcal R}}
\newcommand{\spc}{{\,}}
\newcommand{\sep}{{\qquad}}
\newcommand{\im}{\mathrm{i}}
\newcommand{\id}{\mathbb{I}}
\newcommand{\extN}{N}
\renewcommand{\Im}{\operatorname{Im}}
\renewcommand{\Re}{\operatorname{Re}}
\DeclareMathOperator{\Hodge}{\star}
\newcommand{\pd}{\partial}
\newcommand{\nv}{n_\mathrm{v}}
\newcommand{\CY}{{M_\text{CY}}}
\newcommand{\KD}{\mathcal{D}}
\newcommand{\HM}{\check{\mathcal{M}}}
\newcommand{\hol}{\text{hol}}
\newcommand{\cN}{\mathcal{N}}
\newcommand{\K}{K}

\preprint{IFIC/10-53\\ WITS-CTP-063\\ ITP-UU-10/47\\ SPIN-10/40\\ IFT-UAM/CSIC-10-105}

\title{First-order flows and stabilisation equations for non-BPS extremal black holes}

\author{Pietro Galli\\
Departament de F\'{i}sica Te\`{o}rica and IFIC (CSIC-UVEG)\\
Universitat de Val\`{e}ncia\\
C/ Dr.~Moliner, 50, 46100 Burjassot (Val\`{e}ncia), Spain\\
\email{Pietro.Galli [at] ific.uv.es}}

\author{Kevin Goldstein\\
National Institute for Theoretical Physics (NITHeP)\\
School of Physics and Centre for Theoretical Physics\\
University of the Witwatersrand\\
WITS 2050, Johannesburg, South Africa\\
\email{Kevin [at] neo.phys.wits.ac.za}}

\author{Stefanos Katmadas\\
Institute for Theoretical Physics and Spinoza Institute\\ Universiteit Utrecht\\
Postbus 80.195, 3508 TD Utrecht, The Netherlands\\
\email{s.katmadas [at] uu.nl}}

\author{Jan Perz\\
Instituto de F\'{\i}sica Te\'{o}rica UAM/CSIC\\
C/ Nicol\'{a}s Cabrera, 13--15, C.U. Cantoblanco, 28049 Madrid, Spain\\
\email{jan.perz [at] uam.es}}

\abstract{We derive a generalised form of flow equations for extremal static and rotating non-BPS black holes in four-dimensional ungauged $\extN = 2$ supergravity coupled to vector multiplets. For particular charge vectors, we give stabilisation equations for the scalars, analogous to the BPS case, describing full known solutions. Based on this, we propose a generic ansatz for the stabilisation equations, which surprisingly includes ratios of harmonic functions.}


\begin{document}

\section{Introduction}

The study of black holes in theories with eight or more supercharges resulting from string theory compactifications has proved to be a very useful tool in uncovering some of the structure of the underlying statistical systems. For supersymmetric black holes this task is facilitated by the fact that they exhibit  the attractor mechanism and full supersymmetry enhancement near the event horizon \cite{Ferrara:1995ih, Strominger:1996kf, Ferrara:1996dd}. Using the constraints imposed by supersymmetry, general stationary asymptotically flat solutions have been found in ungauged $\extN = 2$ Einstein--Maxwell supergravity, including higher-derivative corrections, both in four and five dimensions \cite{Behrndt:1997ny, Denef:2000nb, Cardoso:2000qm, Gauntlett:2002nw, Gauntlett:2004qy, Castro:2008ne}. The spatial profile of scalars in these solutions follows a first-order gradient flow, which is integrable to (non-differential) stabilisation equations, expressing the scalars in terms of harmonic functions. On the event horizon (the endpoint of the flow), the values of scalars are dictated by the charges through the attractor equations, independently of the asymptotic boundary conditions (the beginning of the flow).\footnote{Some authors interchange the meaning of the terms ``stabilisation equations'' and ``attractor equations''.}

In contrast, when the requirement that the solutions must preserve some supersymmetry is abandoned, much less is known about the general structure of the supergravity solutions and the microscopic theory behind them. The simplest generalisation of BPS black holes to consider are the extremal black holes which do not preserve any supersymmetry (see \cite{Khuri:1995xq}). These are known to share some desirable features with the BPS branch, most importantly the attractor phenomenon \cite{Ferrara:1997tw,Sen:2005wa,Goldstein:2005hq}.

For theories with $8$ supercharges coupled to vector multiplets in four and five dimensions,\footnote{Since the two are related by dimensional reduction, we do not make a distinction between them in this introduction.} the general structure of these non-BPS extremal solutions is unclear, since only partial results are known. In the static case, a restricted set of examples can be found by simply changing the sign of a subset of the charges, which breaks supersymmetry \cite{Tripathy:2005qp,Kallosh:2006ib}. It was found that the non-BPS solutions exhibit flat directions in the scalar sector, in the sense that the scalars are not completely fixed at the horizon once the charges are chosen \cite{Tripathy:2005qp}. However, these examples are not generic enough --- they contain one less than the minimum number of parameters required for the most general solution to be derived from them by dualities. A solution that does contain enough parameters is called a seed solution.

For cubic prepotentials, an appropriate seed was found in \cite{Cardoso:2007ky, Gimon:2007mh} and the full duality orbit for the $stu$ model was subsequently derived in \cite{Bellucci:2008sv}. This full example clarifies how the non-supersymmetric solutions differ from their BPS counterparts in more than simply changing the signs of charges. In particular, they have flat directions that are subject to symmetries that act along the full flow, including the horizon \cite{Ferrara:1997uz, Bellucci:2006xz, Ferrara:2007tu}.

If one allows for angular momentum, there are two types of single-centre extremal solutions which display attractor behaviour \cite{Astefanesei:2006dd}. The over-rotating (or ergo) branch are very different from the BPS solutions, as they feature an ergoregion and are continuously connected to the Kerr solution \cite{Rasheed:1995zv, Matos:1996km, Larsen:1999pp}. In contrast, the under-rotating (or ergo-free) black holes have a continuous limit to static charged black holes and seem to be tractable using BPS-inspired techniques. Recently, the single-centre under-rotating seed solution and various multi-centred generalisations were found in \cite{Goldstein:2008fq, Bena:2009ev, Bena:2009en, Bena:2009qv, Bena:2009fi}. In these cases, the nontrivial parameter appearing in the static seed solutions can be viewed as the constant part of a harmonic function describing rotation.

Despite the existence of these known solutions, finding an organising principle for their general structure has proven challenging. The best developed approaches are based on four-dimensional supergravity, where electric-magnetic duality limits the possible structures. One such framework is provided by the timelike dimensional reduction of Breitenlohner, Maison and Gibbons \cite{Breitenlohner:1987dg}, which relates black holes, regardless of supersymmetry (or even extremality), to geodesics on the (augmented) scalar manifold. Given sufficient symmetry on the scalar manifold, solutions, including multi-centre black holes, may be generated with powerful group-theoretical methods, cf.~\cite{Gunaydin:2005mx, Gaiotto:2007ag, Bergshoeff:2008be, Bossard:2009at, Bossard:2009my, Bossard:2010mv} and references therein. Unfortunately, this comes at the expense of the results being expressed less explicitly.

A more direct perspective has been offered by the fake superpotential approach of Ceresole and Dall'Agata \cite{Ceresole:2007wx}. They noticed that the rewriting of the effective black hole potential for the scalars \cite{Ferrara:1997tw} as a sum of squares is not unique, leading to more than one type of first-order flow for the scalar fields. The flow, which in the supersymmetric case is governed by the absolute value of the central charge, may be more generally controlled by a different function, called the fake superpotential. The derivation of first-order equations based on a superpotential has been subsequently extended to static non-extremal black holes and for a number of models superpotentials have been identified explicitly \cite{Andrianopoli:2007gt, Perz:2008kh, Ceresole:2009iy, Bossard:2009we, Ceresole:2009vp} (see \cite{Ceresole:2010hq} for a synopsis of these developments and \cite{Miller:2006ay} for earlier related work).

The superpotential method has been first applied to multi-centre black holes in \cite{Galli:2009bj}, which directly generalised \cite{Denef:2000nb}. However, simplifying assumptions restricted the non-super\-symmetric solutions, as in \cite{Gaiotto:2007ag}, to threshold-bound configurations with mutually local charges and unconstrained relative positions of the centres. In view of the recent results on the integrability of the scalar equations of motion in black hole backgrounds \cite{Andrianopoli:2009je, Chemissany:2010zp}, one might expect also the more complicated multi-centre solutions mentioned earlier to be derivable from first-order flows integrated to stabilisation equations.

As a step in this direction, we study extremal under-rotating (ergo-free) black holes in compactifications of Type IIB string theory on Calabi--Yau manifolds, using the formalism of \cite{Denef:2000nb}. In section \ref{eqns-from-action} we relax the additional conditions of \cite{Galli:2009bj} to arrive at the general form of first-order flow equations for stationary extremal black holes. Unfortunately, we find that unlike their previously known special cases, to which they correctly reduce under the relevant assumptions, they generically do not lend themselves to explicit integration.

In section \ref{stab-ansatz}, therefore, we follow a bottom-up approach, trying to find stabilisation equations by rewriting known solutions (expressed in terms of physical scalars or affine coordinates) in a symplectically covariant way, using the projective (homogeneous) coordinates. We find that this is indeed possible for the under-rotating seed solution of \cite{Bena:2009ev}, if one adds a ratio of harmonic functions to the standard vector of harmonic functions appearing in the stabilisation equations. Motivated by this, we introduce an ansatz for the general case that can incorporate all known extremal solutions. Our arguments are independent of the considerations in section \ref{eqns-from-action}, but the general form of the proposed ansatz is compatible with the generic first-order flow equations. However, it is difficult to fully impose it in the rotating case.

Finally, in section \ref{static-potential} we combine our general flow equations with the ansatz in the static case and connect to the fake superpotential formalism. Section \ref{Discussion} is devoted to concluding remarks, whereas in the Appendix we present a general heuristic argument justifying the presence of ratios of harmonic functions in the stabilisation equations.

\section{Bosonic action and special geometry}\label{act_geom}

In rewriting of the effective action as a sum of squares and deriving the flow equations for stationary black holes in $4$-dimensional $\extN=2$ supergravity, we largely follow the method and the notational conventions of the two papers \cite{Denef:2000nb, Galli:2009bj} whose results we generalise (where we also refer the reader for more details and  additional references).

Omitting the hypermultiplets, which are immaterial for our discussion, the relevant bosonic action \cite{deWit:1984pk, deWit:1984px}
\begin{equation}
\label{Ssugra4D}
\Sa_\text{4D}=\frac{1}{16\pi}\int_{M_4} \Bigl(R\Hodge 1 - 2\,g_{a\bar{b}}\,\de z^a\wedge\Hodge\de\bar{z}^{\bar{b}} - \tfrac12 F^I\wedge G_I\Bigr),
\end{equation}
contains  neutral complex scalars $z^a$ (belonging to the $\nv$ vector multiplets) and abelian gauge fields (from both the gravity multiplet and the vector multiplets), all coupled to gravity.

The scalars $z^a$ are affine coordinates on a special K\"ahler manifold, whose metric can be calculated from the K\"ahler potential $\K$: $g_{a\bar{b}} = \partial_a\bar{\partial}_{\bar{b}}\K$, where $\pd_a$ is shorthand for $\pd/\pd z^a$.

The field strengths are defined as $F^I=\de A^I$, where $A^I=(A^0,A^a)$, $a=1,\dotsc, \nv$. The dual field strengths $G_{I}$ are given in terms of the field strengths and the kinetic matrix $\cN_{IJ}$ by
\begin{equation}
G_I=\Im\cN_{IJ}\,\Hodge F^J+\Re\cN_{IJ}\,F^J\spc.
\end{equation}

We will not need the explicit formulae in what follows, but both the K\"ahler potential and the kinetic matrix $\cN_{IJ}$ for the vector fields are derivable from a prepotential, $F$, which is a homogeneous function of degree $2$. The prepotential itself is typically displayed in homogeneous projective coordinates $X^I$ ($z^a = X^a/X^0$) and we will take it to be of the cubic type:
\begin{equation}\label{prep-def}
F=-\frac{1}{6}D_{abc}\frac{X^a X^b X^c}{X^0}=:(X^0)^2 f(z)\spc,\sep
f(z) = -\frac{1}{6}D_{abc} z^a z^b z^c\spc.
\end{equation}

Surface integrals surrounding the sources of the field strengths and their duals define physical magnetic and electric charges, $p^I$ and $q_I$, respectively:
\begin{equation}
p^I = \frac{1}{4\pi}\int_{S^2}F^I\spc, \sep q_I = \frac{1}{4\pi}\int_{S^2}G_I\spc.
\end{equation}

From a geometrical point of view, the above theory can be regarded as the bosonic massless sector of type IIB superstring theory in $10$ dimensions compactified on a Calabi--Yau three-fold $\CY$.\footnote{Thanks to mirror symmetry, one can equivalently use the type IIA picture.} The scalars of the vector multiplets parametrise the moduli space of complex structure deformations of $\CY$. The complex dimension of this scalar manifold is given by one of the Hodge numbers of $\CY$, $\nv = h^{2,1}$, with the K\"ahler potential $\K(z,\bar{z})$ being determined by the unique (up to rescaling), nowhere vanishing holomorphic $(3,0)$-form $\Omega_\hol$, characterising $\CY$:
\begin{equation}
 \K=-\ln\Bigl(\im\int_\CY\Omega_\hol\wedge\bar\Omega_\hol\Bigr).
\end{equation}
It will be more convenient later to work with the \emph{covariantly} holomorphic version of the top form
\begin{equation}
 \label{perioVectorNormalised}
\Omega(z,\bar{z})=\ee^{\K(z,\bar{z})/2}\Omega_{\hol}\spc,
\end{equation}
whose K\"ahler covariant derivative reads
\begin{equation}
\KD\Omega = (\de + \im\,Q)\Omega\spc,
\end{equation}
where $Q = \Im(\partial_a\K\de z^a)$ plays the role of the connection. In components:
\begin{equation}
 \label{covariantDerivativeOmega}
\KD_a\Omega = \partial_a \Omega + \tfrac{1}{2} \partial_a\K\,\Omega\spc, \sep
\bar{\KD}_{\bar{a}}\Omega =
\bar{\partial}_{\bar{a}} \Omega - \tfrac{1}{2} \bar{\partial}_{\bar{a}}\K\,\Omega = 0\spc.
\end{equation}
In the canonical symplectic basis $\{\alpha_I,\beta^J\}$ for the third integral cohomology $H^3(\CY,\mathbb{Z})$, we can expand $\Omega$ as:
\begin{equation} \label{omega-expan}
\Omega = X^I\alpha_I - F_I\beta^I\spc,
\end{equation}
where the coefficients are the periods of the Calabi--Yau manifold with respect to the dual homology basis of three-cycles $\{A^I,B_J\}$:
\begin{equation}
X^I = \int_{A^I}\Omega = \int_\CY\Omega\wedge\beta^I\spc, \sep
F_I = \int_{B_I}\Omega = \int_\CY\Omega\wedge\alpha_I\spc.
\end{equation}
$F_I$ are further identified with the derivatives of the prepotential $F$ with respect to $X^I$: $F_I = \partial F/\partial X^I$ (we hope that no confusion with the gauge field strength two-form $F^I$ arises).

Similarly, the five-form field strength $\mathcal{F}$ of the IIB theory, assumed to take values in $\Omega^2(M_4)\otimes H^3(\CY,\mathbb{Z})$, where $\Omega^2(M_4)$ represents the space of two-forms on spacetime, can be written as
\begin{equation}\label{totalF}
\mathcal{F}= F^I\otimes\alpha_I - G_I\otimes\beta^I\spc.
\end{equation}
By integrating the field strength over an appropriate two-sphere in space, we recover the charges as the coefficients of the three-form $\Gamma\in H^3(\CY,\mathbb{Z})$:
\begin{equation}\label{gamma-def}
\Gamma = \frac{1}{4\pi}\int_{S^2}\mathcal{F} = p^I\alpha_I - q_I\beta^I\spc.
\end{equation}

The five-form $\mathcal{F}$ is self-dual in $10$ dimensions, $\Hodge_{10}\mathcal{F}=(\Hodge\otimes\,\diamond)\mathcal{F}=\mathcal{F}$, where $\Hodge$ and $\diamond$ represent the Hodge operators in, respectively, spacetime and the internal CY manifold $\CY$. A representation of the Hodge operator on the basis forms $\{\alpha_I,\beta^J\}$ can be given in terms of a scalar-dependent matrix $\HM(\mathcal{N})$:
\begin{equation}\label{W-scal}
\begin{pmatrix}\diamond\,\beta^I\\ \diamond\,\alpha_J\end{pmatrix} = \HM^{-1}(\mathcal{N})\begin{pmatrix}\beta^I\\ \alpha_J\end{pmatrix},
\end{equation}
so that the selfduality constraint on $\mathcal{F}$ can be expressed in terms of components as
\begin{equation}\label{W-def}
\HM(\mathcal{N})\begin{pmatrix}\Hodge F^I\\ \Hodge G_J\end{pmatrix}
= \begin{pmatrix}F^I\\ G_J\end{pmatrix}.
\end{equation}

Instead of the canonical basis, one may use the Dolbeault cohomology basis furnished by $\{\Omega,\KD_a\Omega,\bar\KD_{\bar{a}}\bar\Omega,\bar\Omega\}$, diagonalising the Hodge operator $\diam$ on $\CY$:
\begin{equation}\label{CY-hodge}
\diam\Omega=-\im\Omega\spc,\sep
\diam\KD_a\Omega=\im\KD_a\Omega\spc,
\end{equation}
and satisfying
\begin{equation}
\langle\Omega,\bar{\Omega} \rangle = -\im\spc, \sep
\langle\KD_a\Omega,\Dbar{b}{\Omega} \rangle = \im g_{a\bar{b}}\spc, \sep
\langle\KD_a\Omega,\Omega \rangle = 0
\end{equation}
with respect to the antisymmetric intersection product
\begin{equation}\label{inter-prod}
\Iprod{E_1}{E_2}=\int_\CY E_1\wedge E_2\spc.
\end{equation}

In this notation, the central charge $Z$ can be written as
\begin{equation}
Z(\Gamma) = \Iprod{\Gamma}{\Omega} = q_I X^I-p^I F_I\spc,
\end{equation}
and, conversely, one can prove that
\begin{equation}\label{inter-prod-Z}
\Iprod{\Gamma_1}{\Gamma_2} = 2 \Im[- Z(\Gamma_1)\,\bar{Z}(\Gamma_2) + g^{a\bar{b}}\,\KD_a Z(\Gamma_1) \,\bar{\KD}_{\bar{b}} \bar{Z}(\Gamma_2)]\spc,
\end{equation}
where $g^{a\bar{b}}$ is the inverse matrix of $g_{a\bar{b}}$.

Another useful object is the symmetric Hodge product $\Iprod{E_1}{\diamond E_2}$, which introduces a norm on $H^3(\CY,\mathbb{R})$:
\begin{equation}
\lVert E \rVert^2 = \Iprod{E}{\diamond E}\spc.
\end{equation}
An example of its utility in the context of the attractor mechanism is provided by the black hole (or effective) potential
\begin{equation}\label{bh-pot}
V_\BH = \tfrac12 \lVert\Gamma\rVert^2 =
-\tfrac12\begin{pmatrix}p & q\end{pmatrix}\mathcal{M}(\mathcal{N})\begin{pmatrix}p \\ q\end{pmatrix},
\end{equation}
where we suppressed the indices on the charges. (In what follows we will often identify the elements of $H^3(\CY)$ with the associated vectors built out of the components in a symplectic basis, such as $\Gamma$ and $(p^I,q_J)^\mathrm{T}$ here.) The matrices $\mathcal{M}(\mathcal{N})$ and $\HM(\mathcal{N})$ are related to each other by the symplectic metric $\mathcal{I}$,
\begin{equation}
 \HM = \mathcal{I}\mathcal{M}\spc, \sep
\mathcal{I}=\begin{pmatrix}0 & -\mathbb{I}\\ \mathbb{I} & 0\end{pmatrix}, \sep \HM^{-1} = -\HM\spc,
\end{equation}
and are functions of the kinetic matrix $\mathcal{N}_{IJ}$ (see eg.~\cite{Ceresole:1995ca}), although detailed expressions will not be needed in our considerations.

\section{Flow equations from the action}\label{eqns-from-action}

In this section we derive generalised flow equations for non-BPS extremal black holes in Type IIB compactifications on Calabi--Yau manifolds, using the formalism of \cite{Denef:2000nb}. As these equations are not directly integrable, we describe an algorithm for solving them.

\subsection{The action as a sum of squares}\label{sum-of-squares}

Since we are interested in asymptotically flat, stationary extremal black hole configurations, the ansatz that we use for the spacetime metric is:
\begin{equation}
 \label{metricMltc}
 ds^2=-\ee^{2U}(\de t+\omega_i\de x^i)^2+\ee^{-2U}\delta_{ij}\de x^i\de x^j\spc ,
\end{equation}
with the condition $U(x^i),\,\omega_i(x^j)\to 0$ as $r = \sqrt{\delta_{ij} x^i x^j} \to \infty$. The one-form $\omega=\omega_i\de x^i$ encodes the angular momentum of the system.

The action \eqref{Ssugra4D} is not invariant under electromagnetic duality rotations, but as remarked in \cite{Denef:2000nb}, at the cost of discarding manifest Lorentz invariance, one can have a duality invariant formalism \cite{Henneaux:1988gg,Bekaert:1998yp}. For this purpose it is convenient to introduce the following product of spatial 2-forms $\boldsymbol{\mathcal{B}},\boldsymbol{\mathcal{C}} \in \Omega^2(\mathbb{R}^3)\otimes H^3(\CY)$:
\begin{equation}
 \label{Prod2forms}
 (\boldsymbol{\mathcal{B}},\boldsymbol{\mathcal{C}}) = \frac{\ee^{2U}}{1-w^2} \int_\CY \boldsymbol{\mathcal{B}}\wedge\big[
 \boldsymbol{\Hodge_0}(\diam \boldsymbol{\mathcal{C}}) -
 \boldsymbol{\Hodge_0}(w\wedge\diam\boldsymbol{\mathcal{C}})\,w
 +\boldsymbol{\Hodge_0}(w\wedge\boldsymbol{\Hodge_0}\boldsymbol{\mathcal{C}})\big]\spc,
\end{equation}
where by $\boldsymbol{\Hodge_0}$ we denote the Hodge dual with respect to the flat three-dimensional metric $\delta_{ij}$ and define $w=\ee^{2U}\omega$. In general, boldface symbols will be reserved for quantities in the three spatial dimensions.

With this notation, the effective action \eqref{Ssugra4D} can be written as
\begin{equation}
\label{S4Dmltc}
\Sa_{\text{4D eff}}=-\frac{1}{16\pi}\int\de t \int_{\mathbb{R}^3}\Big[2\db U\wedge\boldsymbol{\Hodge}_\mathbf{0}\db U - \tfrac{1}{2}\ee^{4U}\db\omega\wedge\boldsymbol{\Hodge}_\mathbf{0}\db\omega
+2g_{a\bar{b}}\,\db z^a\wedge\boldsymbol{\Hodge}_\mathbf{0}\db \bar{z}^{\bar{b}}+(\boldsymbol{\mathcal{F}},\boldsymbol{\mathcal{F}})\Big].
\end{equation}
As shown in \cite{Denef:2000nb}, this action can be re-expressed as a sum of squares, giving first-order flow equations for stationary supersymmetric black holes, including multi-centre composites:
\begin{align}
&\boldsymbol{\mathcal{F}}-2\I\boldsymbol{\Hodge}_\mathbf{0}\mathbf{D}(\ee^{-U}\ee^{-\im\betad} \Omega )+2\R\mathbf{D}(\ee^{U}\ee^{-\im\betad}\Omega\omega)=0\spc,\label{BPS1}\\
&\mathbf{Q}+\db\betad+\tfrac{1}{2}\ee^{2U}\boldsymbol{\Hodge_0}\db\omega=0\spc,\label{BPS2}
\end{align}
where
\begin{equation}
\label{defDQ}
  \mathbf{D}=\db+\im(\mathbf{Q}+\db\betad+\tfrac{1}{2}\ee^{2U}\boldsymbol{\Hodge_0}\db\omega)\spc,\sep\mathbf{Q}=\I\,(\partial_a\K\db
  z^a)\spc.
\end{equation}

In the light of the considerations in \cite{Ceresole:2007wx}, one might expect similar equations, involving a modified field strength $\boldsymbol{\Ftilde}$ in place of the actual field strength $\boldsymbol{\mathcal{F}}$, to exist for non-supersymmetric extremal black holes as well. In \cite{Galli:2009bj} this was shown to be true for the special case when the fake field strength is related to the real field strength by a constant symplectic matrix. Building on this, we have found a more general way of writing the action as a sum of squares resulting in non-BPS first-order flow equations, based on a non-closed $\boldsymbol{\Ftilde}$ and a new auxiliary one-form $\boldeta$ related to the non-closure of $\boldsymbol{\Ftilde}$. These two objects are constrained by two new equations which need to be satisfied in addition to the flow equations we obtained.

While they will be derived and explained below, for ease of comparison with (\ref{BPS1}, \ref{BPS2}), we state our non-BPS equations now. The first two equations are very similar to the BPS ones:
\begin{align}
\label{G=0} &\boldsymbol{\Ftilde}-2\I\boldsymbol{\Hodge}_\mathbf{0}\mathbf{D}(\ee^{-U}\ee^{-\im\betad} \Omega )+2\R\mathbf{D}(\ee^{U}\ee^{-\im\betad}\Omega\omega)=0\spc,\\
\label{Q+beta+eta+domega}&\mathbf{Q}+\db\betad+\boldsymbol{\eta}+\tfrac{1}{2}\ee^{2U}\boldsymbol{\Hodge_0}\db\omega=0\spc,
\end{align}
with $\boldsymbol{\Ftilde}$ replacing $\boldsymbol{\mathcal{F}}$ in the first equation and $\boldeta$ shifting the second.\footnote{$\mathbf{D}$ and $\mathbf{Q}$ are defined as in (\ref{defDQ}).} In addition, the two equations constraining our two auxiliary variables are:
\begin{align}
\label{constr1}&\bigl(\boldsymbol{\mathcal{F}},\boldsymbol{\mathcal{F}}\bigr)=\bigl(\boldsymbol\Ftilde,\boldsymbol\Ftilde\bigr)+2\db\boldeta\wedge w\spc,\\
\label{constr2}&
\boldeta\wedge\I\Iprod{\boldsymbol{\Gtilde}}{\ee^U\ee^{-\im\betad}\Omega}
  =\Iprod{\db\boldsymbol\Ftilde}{\R(\ee^U\ee^{-\im\betad}\Omega)}\spc-\tfrac12\boldsymbol\db\boldeta\wedge w\spc,
\end{align}
where
\begin{equation}\label{CurlyG}
\boldsymbol{\Gtilde}=\boldsymbol{\Ftilde}-2\I\boldsymbol{\Hodge}_\mathbf{0}\mathbf{D}(\ee^{-U}\ee^{-\im\betad} \Omega )+2\R\mathbf{D}(\ee^{U}\ee^{-\im\betad}\Omega\omega)\spc.
\end{equation}

Now onto the derivation. In \cite{Denef:2000nb} the crucial step in obtaining the first-order manifestly duality-invariant flow equations, solving second-order equations of motion, is to appropriately pair the derivatives of the scalars with the gauge fields and use the invariant product \eqref{Prod2forms} to re-express the Lagrangian. It was found in \cite{Denef:2000nb} that a good choice is
\begin{equation}\label{DenefG}
\boldsymbol{\mathcal{G}}=\boldsymbol{\mathcal{F}}-2\I\boldsymbol{\Hodge}_\mathbf{0}\mathbf{D}(\ee^{-U}\ee^{-\im\betad} \Omega )+2\R\mathbf{D}(\ee^{U}\ee^{-\im\betad}\Omega\omega)\spc.
\end{equation}

Similarly to \cite{Galli:2009bj}, we generalise the above to eq.~\eqref{CurlyG} by replacing the actual field strength with a two-form valued `fake' field strength, $\boldsymbol{\Ftilde}\in \Omega^2(\mathbb{R}^3)\otimes H^3(\CY)$, but here we define it by demanding only that it reproduces the original electromagnetic part of the action
\begin{equation}\label{Ftilde}
\bigl(\boldsymbol{\mathcal{F}},\boldsymbol{\mathcal{F}}\bigr)=\bigl(\boldsymbol\Ftilde,\boldsymbol\Ftilde\bigr)-\boldsymbol\varXi\spc,
\end{equation}
up to a possible extra term described by the three-form $\boldsymbol\varXi$. The form of $\boldsymbol\varXi$ will be determined at the end of this subsection by consistency arguments. Unlike the real field strength, we do not require $\boldsymbol{\Ftilde}$ to be closed.

One can then rewrite the Lagrangian in terms of $\boldsymbol{\Gtilde}$ as
\begin{equation}\label{Lmltc}
\begin{split}
 \mathcal{L}={}&(\boldsymbol{\Gtilde},\boldsymbol{\Gtilde})-4\,(\mathbf{Q}+\db\betad+\boldsymbol{\eta}+\tfrac{1}{2}\ee^{2U}\boldsymbol{\Hodge_0}\db\omega)\wedge\I\langle\boldsymbol{\Gtilde},\ee^U \ee^{-\im\betad}\Omega\rangle\\
&+\db\,[\,2w\wedge(\mathbf{Q}+\db\betad)+4\R\langle\boldsymbol{\Ftilde},\ee^U \ee^{-\im\betad}\Omega\rangle\,]\spc,
\end{split}
\end{equation}
provided that the new one-form $\boldeta$, which needs to be introduced due to the possible non-closure of $\boldsymbol{\Ftilde}$, satisfies
\begin{equation}
  \label{eta}
\boldeta\wedge\I\Iprod{\boldsymbol{\Gtilde}}{\ee^U\ee^{-\im\betad}\Omega}
  =\Iprod{\db\boldsymbol\Ftilde}{\R(\ee^U\ee^{-\im\betad}\Omega)}\spc+\tfrac14\boldsymbol\varXi
  \spc.
\end{equation}
Because adding a total derivative to the Lagrangian does not change the equations of motion, one finds that equation \eqref{eta} needs to hold only up to a total derivative. Observe that the phase $\betad=\betad(\mathbf{x})$ is a priori an arbitrary function.

A sufficient condition for a stationary point of the action (hence, for the equations of motion to be satisfied) is met by neglecting the boundary terms and requiring that the variations of the first two terms in \eqref{Lmltc} vanish separately, which leads to (\ref{G=0}, \ref{Q+beta+eta+domega}). From (\ref{Q+beta+eta+domega}) we obtain $\mathbf{D}=\db-\im\boldeta$, which substituted into (\ref{G=0}) gives:
\begin{equation}\label{EqmScalar_DefFtilde}
 \boldsymbol{\Ftilde}-2\I\bigl[\boldsymbol{\Hodge}_\mathbf{0}(\db-\im\boldeta)(\ee^{-U} \ee^{-\im\betad} \Omega )\bigr]+2\R\bigl[(\db-\im\boldeta)(\ee^{U}\ee^{-\im\betad}\Omega\omega)\bigr]=0\spc.
\end{equation}
Differentiating one finds:
\begin{equation}\label{EqmDifferential=0}
\db\boldsymbol{\Hodge}_\mathbf{0}\db\I\Omt-\db\left(\boldsymbol{\Hodge_0}\boldeta \R\Omt\right)
         - \db\left(\boldeta\wedge w\I\Omt\right) =\tfrac{1}{2}\db\boldsymbol{\Ftilde}\spc,
\end{equation}
where $\Omt=\ee^{-U}\ee^{-\im\betad}\Omega$.\footnote{Note that in \cite{Denef:2000nb} the hat symbolised what we call $\diamond$ here.} We see in particular that, as mentioned earlier, the fake field strength is not necessarily closed.

It is now possible to derive $\boldsymbol\varXi$ in terms of the other quantities appearing in the rewriting. The fundamental observation is that in the dynamical system we are considering, the electromagnetic part of the Lagrangian acts as a potential for the remaining fields and, for solutions of the equations of motion with vanishing action, it is expected to equal the kinetic energy. So, by expressing $\boldsymbol\Ftilde$ through equation \eqref{EqmScalar_DefFtilde}, one can compute $\bigl(\boldsymbol{\tilde{\mathcal{F}}},\boldsymbol{\tilde{\mathcal{F}}}\bigr)$ and find:
\begin{equation}\label{FFrel}\begin{split}
 \bigl(\boldsymbol{\tilde{\mathcal{F}}},\boldsymbol{\tilde{\mathcal{F}}}\bigr)={}&2\db U\wedge\boldsymbol{\Hodge}_\mathbf{0}\db U - \tfrac{1}{2} \ee^{4U}\db\omega\wedge\boldsymbol{\Hodge}_\mathbf{0}\db\omega+2g_{a\bar{b}}\db z^a\wedge\boldsymbol{\Hodge}_\mathbf{0}\db z^{\bar{b}}\\
 &+\ee^{2U}\db w\wedge\boldsymbol{\Hodge}_\mathbf{0}\db\omega+2\db\mathbf{Q}\wedge w\\
={}&\bigl(\boldsymbol{\mathcal{F}},\boldsymbol{\mathcal{F}}\bigr)+\ee^{2U}\db w\wedge\boldsymbol{\Hodge}_\mathbf{0}\db\omega+2\db\mathbf{Q}\wedge w\spc,
\end{split}
\end{equation}
from which, using \eqref{Ftilde}, \eqref{Q+beta+eta+domega} and integration by parts, one obtains
\begin{equation}
\label{Xi}
 \boldsymbol\varXi=-2\db\boldeta\wedge w\spc.
\end{equation}
Finally, substituting (\ref{Xi}) in (\ref{Ftilde}) and (\ref{eta}) leads to (\ref{constr1}, \ref{constr2}).

In summary, we have obtained a non-supersymmetric generalisation (\ref{G=0}, \ref{Q+beta+eta+domega}) of first-order equations \cite{Denef:2000nb}. The generalised equations are expressed in terms of a fake field strength $\boldsymbol{\Ftilde}$, constrained by \eqref{Ftilde} to reproduce the original gauge part of the action. The non-closure of $\boldsymbol{\Ftilde}$ necessitates the introduction of a new, compensating object, $\boldeta$, in eq.~\eqref{eta}. The auxiliary three-form $\boldsymbol\varXi$ appearing in \eqref{Ftilde} and \eqref{eta} can be expressed in terms of other quantities through eq.~\eqref{Xi}. In comparison with the supersymmetric case we thus have two more unknowns, $\boldsymbol{\Ftilde}$ and $\boldeta$, constrained by (\ref{constr1}, \ref{constr2}). Since they are mutually related, in any model for which $\boldsymbol{\Ftilde}$ can be obtained by other means (as in section \ref{superpotential}), $\boldeta$ can be eliminated as well.

\subsection{Solving the equations}\label{sol-by-ans}

Whenever $\db\boldsymbol{\Ftilde} = 0$ and $\boldsymbol{\eta} = 0$, equation \eqref{EqmDifferential=0} reduces to the Laplace equation
\begin{equation}
2\,\db\boldsymbol{\Hodge}_\mathbf{0}\db\I\Omt = 0\spc,
\end{equation}
which can be integrated to so-called stabilisation equations:
\begin{equation}\label{stabilisation-harmonic}
2\I(\ee^{-U}\ee^{-\im\betad} \Omega)=\cH\spc.
\end{equation}
These express the period vector in terms of (possibly multi-centred) harmonic functions $\cH$ throughout the flow.

Using this result back in \eqref{EqmScalar_DefFtilde} one finds that the fake field strength is given by
\begin{equation}\label{gauge-BPS}
\boldsymbol{\Ftilde}=\boldsymbol{\Hodge_0}\db \cH -2\,\db(\ee^{2U}\R\Omt\omega) \spc.
\end{equation}
This is manifestly true for the supersymmetric case \cite{Behrndt:1997ny, Denef:2000nb} and for the non-BPS setting considered in \cite{Galli:2009bj}, for electric $(p^0, 0; 0, q_a)$ or magnetic $(0, p^a; q_0, 0)$ charge configuration and axions $\Re z^a$ set to zero.\footnote{By charge redefinitions one can generate physically equivalent solutions also for other charge configurations \cite{Nampuri:2010um}.} In the BPS case, one has $\boldsymbol{\Ftilde} = \boldsymbol{\mathcal{F}}$ by assumption, so that the vector of electric and magnetic charges is determined through \eqref{gamma-def} to be
\begin{equation}
\Gamma = \frac1{4\pi}\,\int_{S^2} \boldsymbol{\Hodge_0}\db \cH\spc ,
\end{equation}
or in other words equal to the poles of the harmonic functions $\cH$. For non-supersymmetric solutions, $\boldsymbol{\Ftilde}$ is related to $\boldsymbol{\mathcal{F}}$ by charge sign reversals, and the same holds for the poles of the harmonic functions $\cH$ in \eqref{gauge-BPS} compared to the physical charges.

In its general form, however, equation \eqref{EqmDifferential=0} cannot be solved directly, since the period vector $\Omega$ (scalars), $\boldeta$, $\omega$ and $\boldsymbol\Ftilde$ are all unknown and constrained by \eqref{constr2}. A way out of this problem is to first make an ansatz for $\Im\Omt$, try to solve it for $\Omega$ and $U$, find $\boldeta$ and $\omega$ from \eqref{Q+beta+eta+domega} and then, by using \eqref{EqmScalar_DefFtilde} as a definition for $\boldsymbol\Ftilde$, check if (\ref{constr1}, \ref{constr2}) are satisfied.

Let us see explicitly how to do that: we start by making an ansatz of the type
\begin{equation}
\label{ansatz_rotBH}
2\I(\ee^{-\im\betad}\ee^{-U}\Omega)=\cJ\spc,
\end{equation}
with $\cJ$ a vector containing all the parameters in terms of which the solutions will be expressed. The solution for the components of  $\Omt$  (and hence for scalars) can then be obtained in the same way as the solutions to supersymmetric stabilisation equations (\cite{Shmakova:1996nz}, see also~\cite{Behrndt:2005he}, section 2). We will indicate all quantities calculated with the aid of the ansatz by adding the subscript $\cJ$.

We then proceed differentiating both sides of \eqref{ansatz_rotBH} and subsequently taking the intersection product with the real and imaginary part of $\Omega$ and with $\KD_a\Omega$. With the definitions $\Psi:=-\Iprod{\db \cJ}{\Omega}$ and $\KD_a\Psi:=-\Iprod{\db \cJ}{\KD_a\Omega}$ one obtains:
\begin{align}
\label{dU=Re} \db U&=-\ee^U\R(\ee^{-\im\betad}\Psi)\spc,\\
\label{dbeta+Q=Im} \db\betad+\mathbf{Q}&=\ee^{U}\I(\ee^{-\im\betad}\Psi)=-\tfrac{1}{2}\ee^{2U}\Iprod{\db \cJ}{\cJ}\spc,\\
 \label{dz=DPsi}\db \bar z^{\bar b}&=-g^{a\bar b}\ee^{U}\ee^{-\im\alpha}\KD_a\Psi\spc.
\end{align}
Note that \eqref{dU=Re} and \eqref{dz=DPsi} are the flow equations for the warp factor and the scalars, while \eqref{dbeta+Q=Im} gives an explicit relation between $\betad$ and the other quantities appearing in the rewriting. More specifically, \eqref{dbeta+Q=Im} combined with \eqref{Q+beta+eta+domega} eliminates $\betad$, giving:
\begin{equation}\label{dJ-eta}
 \Iprod{\db \cJ}{\cJ}=2\ee^{-2U}\boldsymbol{\eta}+\boldsymbol{\Hodge_0}\db\omega\spc.
\end{equation}
If we make an ansatz also for the angular momentum of the black hole $\omega_\BH$ (which must be expressed in terms of parameters appearing in $\cJ$), we arrive at an expression for $\boldeta$:
\begin{equation}\label{etaJ}
\boldsymbol{\eta}_{\cJ,\omega}=\tfrac12\ee^{2U_\cJ}\Bigl(\Iprod{\db \cJ}{\cJ}-\boldsymbol{\Hodge_0}\db\omega_\BH\Bigr).
\end{equation}

Independently of the ansatz, we can use \eqref{EqmScalar_DefFtilde} as a definition of $\boldsymbol{\Ftilde}$ and then substitute \eqref{EqmDifferential=0} and \eqref{G=0} in \eqref{constr2}. The left-hand side is clearly zero, whereas on the right-hand side we have an intersection product that we know how to compute. Neglecting the total derivative results in:
\begin{equation}\label{constraint_finalForm}
\Iprod{\db\boldsymbol\Ftilde}{\R(\ee^{2U}\Omt)} =\Iprod{2\,\db\boldsymbol{\Hodge}_\mathbf{0}\db\I\Omt}{\R(\ee^{2U}\Omt)}-\boldeta\wedge\boldsymbol{\Hodge_0}\boldeta-\ee^{2U}\boldeta\wedge\db\omega+\tfrac12\boldeta\wedge\db w\spc.
\end{equation}
The last term here cancels the last term of \eqref{constr2}. This means that once all the variables have been expressed in terms of the parameters in
the vector $\cJ$, the consistency of the ansatz with the first-order equations (\ref{G=0},
\ref{Q+beta+eta+domega}) and the constraint \eqref{constr2} can be verified by checking whether the following equation is satisfied:
\begin{equation}\label{Check}
\ee^{2U_\cJ}\Iprod{\db\boldsymbol{\Hodge}_\mathbf{0}\db \cJ}{\R\Omt_\cJ}=\boldeta_{\cJ,\omega}\wedge\boldsymbol{\Hodge_0}\boldeta_{\cJ,\omega}+\ee^{2U_\cJ}\boldeta_{\cJ,\omega}\wedge\,\db\omega_\BH\spc.
\end{equation}
As this is an equation for $\cJ$, in principle it determines an ansatz that satisfies \eqref{G=0}, \eqref{Q+beta+eta+domega} and \eqref{constr2}, even though in practice one would not solve it for $\cJ$, regarding it instead as a check for the specific form of an ansatz assumed beforehand. One would also still have to ensure that \eqref{constr1} is satisfied, which can be a rather non-trivial task. Finally, not all the parameters in the ansatz may be constrained by the equations of motion but should rather be fixed by appropriate boundary conditions.

In the next section we will discuss known black hole solutions which satisfy (\ref{Check}) with a nontrivial $\boldeta$ and propose a generic ansatz.

\section{Stabilisation equations from an ansatz}\label{stab-ansatz}

An important result for BPS black holes is the direct integrability of the first order flow equations to stabilisation equations, even for multiple centres. As described in the beginning of section \ref{sol-by-ans}, this result can be extended to some non-supersymmetric solutions. These examples, however, are not generic, in the sense that applying dualities on them does not lead to the most general non-BPS solution.

Given the flow equations in section \ref{sum-of-squares}, one expects to find a nontrivial $\boldeta$ and $\boldsymbol\Ftilde$ in the general case. The non-closure of $\boldsymbol\Ftilde$ implies that the corresponding expression for the period vector $\Omega$ in terms of charges and integration constants should be an anharmonic extension of \eqref{stabilisation-harmonic}, which must still be consistent with symplectic reparametrisations. To gain intuition about the possible terms, one can follow a bottom-up approach. Therefore, we consider known explicit solutions and rewrite the physical scalars $z^a$ and the metric in terms of $\Omega$, aiming towards a generic ansatz that covers all single-centre solutions.

\subsection{Special solutions}

In order to be as general as possible and to minimise the ambiguity introduced in the process, we find it illuminating to start with a rotating black hole solution, so that the presence of an extra harmonic function describing angular momentum can provide guidance. Consider the rotating extremal black hole of \cite{Bena:2009ev}, which can be used as the seed solution for four-dimensional under-rotating black holes \cite{Rasheed:1995zv, Matos:1996km, Larsen:1999pp} in theories with cubic prepotentials. This is an almost BPS \cite{Goldstein:2008fq} solution in five-dimensional supergravity described by the harmonic functions:
\begin{equation}\label{har_fun_one}
  H^0=h^0+\frac{p^0}{r}\spc,\sep H_a=h_a+\frac{q_a}{r}\spc,\sep M=b+ \frac{J\,\cos{\theta}}{r^2}\spc,
\end{equation}
where $h^0$, $h_a$, $b$ are constants that are related to asymptotic moduli, $-p^0$ is the Kaluza--Klein magnetic charge, $q_a$ are the electric charges and $J$ is the angular momentum of the solution. Therefore, the associated four-dimensional charge vector is defined by the harmonic functions:
\begin{equation}\label{phys-charge}
  \cH_c=(-H^0,\,0\, ;\,0,\,H_a)\spc,
\end{equation}
whereas $M$ controls the angular momentum and is invariant under symplectic transformations.

Using the 4D/5D dictionary of \cite{Gaiotto:2005gf, Gaiotto:2005xt, Behrndt:2005he}, one can rewrite the full solution given in five-dimensional notation \cite{Bena:2009ev} in terms of variables natural from the four-dimensional perspective. The metric is as in \eqref{metricMltc}, while the resulting expressions for the gauge fields and scalars in our notation are:
\begin{equation}
  \sF=\boldsymbol{\Hodge_0}\db \cH_c -2\,\db(\ee^{2U}\!\R\Omt\,\omega) \spc, \sep
   2\Im\Omt=\cJ\equiv\,\cH+\cR\spc,\label{stab-eqs-02}
\end{equation}
where we use again the shorthand $\Omt=\ee^{-U}\ee^{-\im\betad}\Omega$ as in section \ref{eqns-from-action}. $\cJ$ is written in terms of a harmonic part, $\cH$, and a part containing ratios of harmonic functions $\cR$, which are respectively given by:
\begin{equation}
  \cH=(H^0,\,0\, ;\,0,\,H_a)\spc,\sep \,\cR=\Bigl(0,\,0\, ;\,-\frac M {H^0} ,\,0\Bigr)\spc.\label{h-r-02}
\end{equation}
Finally, the metric functions are given by:
\begin{equation}\label{scale-fact}
  \boldsymbol{\Hodge}_\mathbf{0}\db {\omega} = \db M \spc, \sep
  \ee^{-2U}=\im\Iprod{\Omt}{\bar{\hat{\Omega}}}=\sqrt{I_4(\mathcal{H}) -M^2}\spc.\\
\end{equation}
Here, $I_4$ is the quartic invariant that appears in the entropy formula for cubic prepotentials (see \cite{Cerchiai:2009pi} for explicit expressions) and the physical scalars are given by $z^a=X^a/X^0$, as usual.

The expression for the gauge fields in \eqref{stab-eqs-02} parallels the form of the BPS solutions \eqref{gauge-BPS}, differing in that the vector of harmonic functions associated to the physical charges \eqref{phys-charge}, is related to the one appearing in the scalars by a single sign flip, similar to \cite{Kallosh:2006ib}. The period vector $\Omega$ is again determined through stabilisation equations similar to \eqref{stabilisation-harmonic}, so that the scalars are given in terms of harmonic functions describing the flow from asymptotic infinity to the horizon. In particular, the asymptotic values of the scalars are controlled by the constant parts of the harmonic functions $\cH$ and $M$, whereas the attractor equations, obtained in the limit $r\rightarrow 0$, are controlled by the charges and the angular momentum \cite{Astefanesei:2006dd}.

The novel addition to $\cJ$ is a ratio of harmonic functions that was not present in previous attempts to write non-BPS stabilisation equations and allows for nontrivial axions. Deferring the comparison to the rewriting of section \ref{eqns-from-action} for the next section, we note that this solution leads through \eqref{etaJ} to a nontrivial $\boldeta$, given by
\begin{equation}\label{eta-sol}
\boldsymbol{\eta} =\ee^{2U}\Iprod{\db \cR}{\cH}=-\ee^{2U}\,H^0\,\db \left(\frac{M}{H^0}\right),
\end{equation}
which demonstrates how $\boldeta$ is related to the anharmonic part of the solution.

In the static limit, it is possible to show that all constraints and flow equations of the previous section are indeed satisfied, if $\cJ$ in \eqref{ansatz_rotBH} is identified with the one in \eqref{stab-eqs-02}. The presence of a nontrivial $\boldeta$ that follows from the ratio $M/H^0$ through \eqref{eta-sol} is crucial in this respect. In the rotating case, we have verified the constraint \eqref{Check}, but it is more challenging to verify the flow equations and especially the first constraint \eqref{constr1}.

As this is the seed solution for under-rotating extremal black holes, one can apply duality rotations on the full rotating solution using the stabilisation equations \eqref{stab-eqs-02} to find the most general solution. Imposing that the angular momentum harmonic function $M$ is invariant under duality transformations, the result is that a ratio of harmonic functions is generated in all other cases as well.

For example, in the case of the $stu$ model, one can explicitly dualise to the frame with only two charges present, corresponding to a D0-D6 brane system in Type IIA theory. For this model, the prepotential is as in \eqref{prep-def} with $D_{abc}=|\varepsilon_{abc}|$ and the scalar sector is then described by the choice (no sum on $a=1,2,3$):
\begin{equation}\label{h-r-06}
  \cH=\Bigl(H^0,\,\frac 1 {\lambda^a} H^{a}\, ;\,H_0 ,\,\lambda^a H_{a}\Bigr)\spc,\sep
\cR=\frac18 \frac M {H_0^+}\Bigl(1,\,\frac 1 {\lambda^a}\, ;-1 ,\,-\lambda^a\Bigr)\spc,
\end{equation}
where
 \begin{gather}\label{har_fun_two}
  H_I=h_I+\frac{q_0}{r}\spc,\sep H_0^+= \frac14\Bigl(h_0 + \sum_a h_a\Bigr) +\frac{q_0}{r}\\
    H^I=-\lambda^3 H_I \spc, \sep H^{0\,+}=-\lambda^3 H_0^+ \spc,\sep
   D_{abc} \lambda^a\lambda^b\lambda^c \equiv \lambda^3\spc,\\
  \ee^{-4U}=I_4(\cH) -M^2=(H_0\,H^0)^2 -M^2\spc,
\end{gather}
and $\lambda^3$ must be a constant. Note that the individual constants $\lambda^a$ appear only as multiplicative factors in $\cH$ and $\cR$, but not in $\ee^{-U}$, which depends only on the physical harmonic functions $H_0$ and $H^0$. It follows that the metric and gauge fields depend only on the combination $\lambda^3$, so that two of the $\lambda^a$ correspond to flat directions. The structure in \eqref{h-r-06} is consistent with the results on D0-D6 attractors in \cite{Nampuri:2007gv} and seems to be generic for D0-D6 solutions for all cubic prepotentials.

It is interesting to note that unlike in \eqref{stab-eqs-02}, the harmonic part $\cH$ is not related to the charges by sign flips, as one might expect. In fact the electric solution is special, in the sense that the flat directions can be described through \eqref{stab-eqs-02} by simply allowing for the missing harmonic functions to be constants, at the cost of making $\cR$ more complicated, but still proportional to a single ratio as in the D0-D6 case. On the other hand, for both solutions the angular momentum harmonic function can be invariantly characterised by $M=\Iprod\cH \cR$. The flat directions described by the $\lambda^a$ are zero modes of this equation.

\subsection{The ansatz}\label{the-ansatz}

On account of the above observations, it is natural to propose an ansatz for the period vector that contains harmonic functions and ratios of harmonic functions, disregarding the precise relation to the physical charges, which is to be fixed later. In fact, it is simple to see that imposing consistency of any generic ansatz $\Im \Omega \sim \cH+ \cR$, leads to inverse harmonic functions. Since one can compute $\Iprod\cH \cR$ in two ways:
\begin{equation}\label{moduli-indep}
\Iprod{\cH}{\cR}= \Im \Iprod{\Omt}{\cR}=-\Im \Iprod{\Omt}{\cH}\spc,
\end{equation}
where $\cH$ and $\cR$ are a priori independent, it follows that $\Iprod\cH \cR$ must be a scalar-independent quantity. The only other fields in the system are the scale factor and the rotation form $\omega$ in the metric, but since $\Iprod\cH \cR$ does not carry a scale,\footnote{Here we refer to the symmetry of \eqref{Lmltc} under $\ee^{U}\rightarrow \ee^D \ee^{U}$, $\Omega\rightarrow \ee^D \Omega$, $g_{ij}\rightarrow \ee^{2D} g_{ij}$ for constant $D$, inherited from the full conformal formulation of the theory \cite{Mohaupt:2000mj}.} it cannot depend on  $\ee^{U}$, in accord with the explicit solution above, where $\Iprod\cH \cR=M$. In the static limit $M$ reduces to a constant, in which case the constraint could be solved even if $\cR$ were harmonic, but in the rotating case one has to reproduce the full function $M$, which depends on the angular coordinates. This implies a structure as in \eqref{stab-eqs-02}, with the anharmonic part, which then must be present even when the angular momentum is turned off. A more extensive argument about the kind of zero modes allowed for the scalars, which leads to the same conclusion, is given in the Appendix.

Based on the linearity of symplectic reparametrisations and the fact that \eqref{h-r-02} and \eqref{h-r-06} are seed solutions, we expect the structure seen in the previous section to be universal for all under-rotating extremal black holes. In other words, we take the point of view that there is no essential difference between static non-supersymmetric and under-rotating black holes, since they are continuously connected by setting to zero the nonconstant part of a single harmonic function, as in \eqref{har_fun_one}. Therefore, we propose the following form for the stabilisation equations for the scalars and the angular momentum:
\begin{gather}
  2\Im\Omt \equiv 2\Im(\ee^{-U}\ee^{-\im\betad}\Omega )
   = \cH + \cR \spc,\label{ansatz}\\
  \boldsymbol{\Hodge}_\mathbf{0}\db \omega = \Iprod { \db \cH} \cH
                            + \db \Iprod {\cH}\cR \spc,\label{ansatz2}
\end{gather}
where $\cH$ is a vector of harmonic functions and $\cR$ is a vector of ratios of harmonic functions. The integrability condition of the last equation implies that their symplectic inner product $\Iprod {\cH}\cR$ is a harmonic function, while the scale function of the metric is given by:
\begin{equation}\label{scale-fun}
  \ee^{-2U}=\im\Iprod{\Omt}{\bar{\hat{\Omega}}}=\sqrt{I_4(\cH + \cR)}\spc.
\end{equation}
Note that when $\cR=0$ and the charges carried by $\cH$ are identified with the physical charges, one recovers the BPS stabilisation equations, as required. More generally, for a physically reasonable solution the harmonic and inverse harmonic functions in \eqref{ansatz} are quite restricted due to various consistency constraints, both generic and based on known explicit solutions. The rest of this section is devoted to a discussion of these generic constraints and some of their implications.

A first requirement is that in the near-horizon limit the scale factor $\ee^{-4U}$ of an under-rotating black hole must reduce to \cite{Astefanesei:2006dd}:
\begin{equation}\label{hor-eu}
\ee^{-4U} \propto |I_4(\Gamma)|- J^2\cos^2{\theta}\spc,
\end{equation}
where $I_4(\Gamma)$ is the quartic invariant of the model and $J$ is the angular momentum. In the simple case of vanishing angular momentum, $\cR$ is proportional to inverse harmonic functions and thus vanishes near the horizon. Therefore, a harmonic piece must always be present in the right hand side of \eqref{ansatz}, to make sense of the static solution in the near-horizon region. Similar comments then apply for the full rotating case, hence it is impossible to have a physical solution for the scalars based purely on inverse harmonic functions.

Going over to the constraints posed by the form of the full solution, observe that in the (necessarily static) BPS case the full scale function is simply $\ee^{-4U}=I_4(\cH)$, where the charges are replaced by their corresponding harmonic functions. Similarly, for the $stu$ model, where the most general non-BPS static black hole was explicitly constructed in \cite{Bellucci:2008sv} using the seed solution of \cite{Cardoso:2007ky, Gimon:2007mh}, it has been shown that the scale factor is shifted as $\ee^{-4U}\sim I_4(\cH) - b^2$, where $b$ is a constant that does not depend on the charges.

Interestingly, for the known under-rotating seed solution the expression for $\ee^{-U}$ in \eqref{scale-fact} can again be found from \eqref{hor-eu} by replacing the charges and angular momentum by harmonic functions. Moreover, the additional constant $b$ of \cite{Cardoso:2007ky, Gimon:2007mh, Bellucci:2008sv} is identified with the constant piece in the harmonic function for the angular momentum in \eqref{scale-fact}, as in \cite{Goldstein:2008fq, Bena:2009ev}. Therefore it is reasonable to expect that generically the scale factor is a function of the harmonic functions for the charges and angular momentum, thus allowing for the presence of a possible residual constant in the static solutions, when $J$ is set to zero.

Now, for an ansatz of the type \eqref{ansatz} to describe the known solutions, the vector $\cR$ must be such that \eqref{scale-fun} is consistent with the above comments, in particular with \eqref{scale-fact}, so that
\begin{equation} \label{eu_taylor}
\ee^{-4U} = I_4(\cH + \cR) = I_4(\cH) -  \Iprod \cH \cR ^2 \spc.
\end{equation}
This equality poses very strong restrictions on $\cR$, as it does not appear in linear, cubic or quartic terms. In particular, the components of $\cR$ must be such that $I_4(\cR)$ and its first derivatives vanish, implying that it must have at most as many independent components as a two-charge small black hole. Then, given $\cH$ and a model in which $I_4$ is known, the linear term in $\cR$ should vanish, further restricting its independent components. Indeed, $\cR$ appears to have only one independent component in the explicit solutions \eqref{h-r-02} and \eqref{h-r-06}.

For symmetric cubic models this can be made more precise, by Taylor expanding the left hand side of \eqref{eu_taylor} explicitly. In these models, the quartic invariant can be rewritten in terms of the central charge as
\begin{equation}
    I_4=(i_1-i_2)^2+4 i_4-i_5\spc,
    \label{I4general}
\end{equation}
where
\begin{gather}
\begin{alignat}{2}
i_1 &= Z\bar{Z}\spc, &\mspace{52mu}&
i_2 = g^{a\bar b} \KD_a Z\bar{\KD}_{\bar b}\bar{Z}\spc,
\\
i_3 &= \tfrac{1}{3}\Re( Z N_3(\bar{Z}) )\spc, &&
i_4 = -\tfrac{1}{3}\Im( Z N_3(\bar{Z}) )\spc,
\end{alignat}\label{i4}
\\
i_5 = g^{a\bar a}D_{abc}D_{\bar a \bar b \bar c}
       g^{b\bar d}g^{c\bar e} g^{d\bar b}g^{e\bar c}
        \bar{\KD}_{\bar d}\bar{Z}\,\bar{\KD}_{\bar a}\bar{Z}\,
        \KD_d Z \, \KD_e Z \spc, \label{i5}
\end{gather}
and
\begin{equation}
    N_3(\bar{Z})=D_{abc}g^{a \bar a}g^{b \bar b}g^{c \bar c}
         \bar{\KD}_{\bar a}\bar{Z}\,\bar{\KD}_{\bar b}\bar{Z}\,
         \bar{\KD}_{\bar c}\bar{Z}\spc.
\end{equation}
Although these five invariants all depend on the scalar fields and the charges, the combination in \eqref{I4general} is scalar independent. In this form, it is easy to expand $I_4(\cH+\cR)$ and separately consider the different terms, since $Z$ and its derivatives are linear in the charges. Furthermore, as shown in \cite{Ceresole:2010nm}, there are relations between the invariants above when the charge vector is that of a small black hole. The previous discussion suggests that $\cR$ should have only one independent component, so we assume that it lies in a doubly critical orbit, in which case
\begin{equation}\label{doub-crit}
i_2(\cR)=3i_1(\cR)\spc;\sep i_3(\cR)=0\spc;\sep i_4(\cR)=2i_1^2(\cR)\spc;\sep i_5(\cR)=12 i_1^2(\cR)\spc.
\end{equation}
A straightforward expansion of \eqref{I4general}, using \eqref{doub-crit} leads to
\begin{equation}\label{I4_symmetric}
I_4(\cH + \cR) = I_4(\cH) + \Iprod{\delta I_4(\cH)}{\cR} - \Iprod{\cH}{\cR}^2 \spc,
\end{equation}
where $\delta I_4(\cH)$ denotes the derivative of $I_4(\cH)$ and the identity \eqref{inter-prod-Z} was used. Thus, the quadratic term reorganizes itself in the desired form without further assumptions.\footnote{Conversely, the decomposition in \cite{Ferrara:2010ug} can be used to show that the quartic invariant takes this form only if $\cR$ lies in a doubly critical orbit \cite{Marrani:priv}.\label{orbit}} For a given model, $\cR$ can then be determined by demanding that the linear term vanishes.

This requirement is enough to ensure that the ansatz \eqref{ansatz}, together with the above assumptions, automatically satisfies the constraint \eqref{constr2}, as we now show. First, note that for the ansatz in \eqref{ansatz}, $\boldeta$ takes the form
\begin{equation}\label{eta-ans}
\boldsymbol{\eta} =\ee^{2U}\Iprod{\db \cR}{\cH}\spc,
\end{equation}
as in \eqref{eta-sol}. In section \ref{sol-by-ans} it was shown that the constraint \eqref{constr2} is equivalent to \eqref{Check}, which in view of the last result reads
\begin{equation}\label{Check-ans}
\Iprod{\db\boldsymbol{\Hodge}_\mathbf{0}\db \cJ}{\R\Omt}
=\ee^{2U}\,\Iprod{\db \cR}{\cH} \wedge\Iprod{\boldsymbol{\Hodge_0}\db \cH}{\cR}\spc.
\end{equation}
Using \eqref{scale-fun} and \eqref{eu_taylor}, one can then show that
\begin{equation}\label{eta-con-show}
\Iprod{\db\boldsymbol{\Hodge}_\mathbf{0}\db \cJ}{\R\Omt}=
\tfrac12\, \ee^{2U} \Iprod\cR\cH\,\Iprod{\db\boldsymbol{\Hodge}_\mathbf{0}\db \cR}\cH = -\ee^{2U} \Iprod\cR\cH\,\Iprod{\boldsymbol{\Hodge}_\mathbf{0}\db \cR}{\db\cH}\spc,
\end{equation}
where we used the identity \cite{Bates:2003vx}
\begin{equation}
\R\Omt= \tfrac12\,\begin{pmatrix}\frac{\partial}{\partial \cJ_I}\Bigr. \\ \frac{\partial }{\partial \cJ^J} \end{pmatrix}\ee^{-2U}\spc,
\end{equation}
and the last step follows from the fact that $\Iprod\cR\cH$ is a harmonic function. Finally, since $\cR$ depends only on a single ratio of the form $\Iprod\cH\cR/\bar H$ (with $\bar H$ a harmonic function, cf.~\eqref{h-r-06}), it is possible to show that
\begin{equation}
\Iprod\cH\cR\,\db \cR= -\Iprod{\db\cR}{\cH}\,\cR\spc.
\end{equation}
Combining the last relation with \eqref{eta-con-show}, the constraint \eqref{Check-ans} is identically satisfied, so that (\ref{ansatz}, \ref{ansatz2}) is a solution of the constraint \eqref{constr2}. The seed solution \eqref{stab-eqs-02} also satisfies these relations by construction.

This is a rather nontrivial result, as \eqref{Check}, due to \eqref{etaJ}, is a quartic equation for $\cJ$. Assuming this to be the general solution, the only constraint remaining at this stage is \eqref{constr1}, which generalises the constraint on the fake superpotential for static black holes \cite{Ceresole:2007wx} to the case of under-rotating black holes. However, it is difficult to verify \eqref{constr1} and \eqref{EqmScalar_DefFtilde} explicitly for the seed solution \eqref{stab-eqs-02}, or find the general solution. In the next section, we give a more detailed comparison to the ansatz (\ref{ansatz}, \ref{ansatz2}) in the static limit.

It follows that the only object missing for a complete characterisation of the ansatz for extremal solutions is an explicit form for $\cH$ given a vector of physical charges. In view of the flow equations in the previous section, that would be equivalent to solving \eqref{constr1} which, upon using (\ref{ansatz}, \ref{ansatz2}) to determine the scalars and $\boldeta$, becomes a quadratic equation for the physical charges in terms of $\cH$ and $\cR$.

Unfortunately, solving this constraint is not a straightforward task. The only a priori requirement on $\cH$ is that it must be ``BPS'' in the sense that $I_4(\cH)\!>\!0$ and that its quartic invariant should be related to the one of the physical charges by a sign flip. In fact the result should not be unique, as one can expect in view of the non-uniqueness in the rewriting \eqref{constr1}. A manifestation of this ambiguity is seen in \eqref{h-r-06}, where the two extra unconstrained parameters in $\lambda^a$ represent the flat directions of the scalar sector. On the other hand, the relation between $\cH$ and the physical charges must be the same throughout the flow, as follows from \eqref{ansatz}, so that an attractor analysis would be sufficient for this purpose. In any case, one can always dualise the stabilisation equations for the seed solutions above to find any other solution and we comment on a possible way to construct $\cH$ at the end of the next section.

\section{The static limit}\label{static-potential}

In this section we specialise the results of section \ref{eqns-from-action} to the static case, using the ansatz of section \ref{stab-ansatz}, and connect to the fake superpotential formalism.

\subsection{The static flow equations}\label{static}

The static limit of the results in section \ref{eqns-from-action} leads to several simplifications, since the solutions are necessarily spherically symmetric. This implies that $\omega=0$ and all quantities depend only on the radial coordinate. Similarly as for the actual field strength, spherical symmetry implies that the modified field strength $\boldsymbol\Ftilde$ is of the form:
\begin{equation}
 \boldsymbol{\Ftilde}=\sin\theta\,\de\theta\wedge\de\varphi\otimes\Gamt\spc,
\end{equation}
where now $\Gamt\in H^3(\CY)$ is fibred along $r$. By \eqref{constr1}, it must reproduce the same black hole potential $V_\BH$ as the physical charge $\Gamma$:
\begin{equation}\label{GamtGamt=V}
\tfrac12\lVert\Gamt\rVert^2 = V_\BH = \tfrac12\lVert\Gamma\rVert^2\spc.
\end{equation}

In this setting one chooses the arbitrary function $\ee^{\im\alpha}$ to be the phase of $\Iprod{\Gamt} {\Omega}$. In terms of the inverse radial coordinate $\tau=1/r$ the first-order equations (\ref{G=0}, \ref{Q+beta+eta+domega}) with $\boldeta = \eta\,\de\tau$ reduce to\footnote{The signs depend on the conventions chosen for the Hodge dual.}
\begin{gather}
\label{eqmsimpleSimplified}
2\,\partial_{\tau}\I\bigl(\ee^{-U}\ee^{-\im\alpha}\Omega \bigr)-2\,\eta\,\R\bigl(\ee^{-U}\ee^{-\im\alpha}\Omega\bigr)=-\tilde{\Gamma}\spc,\\
\eta=-\dot\alpha-Q_\tau=-\Im(\Iprod{\dot\Gamt}{\Omega}/Z(\Gamt))\spc,\label{etaS}
\end{gather}
where the second relation turns out to be equivalent to the second constraint \eqref{constr2}. Observe again that the presence of a nontrivial $\eta$ is essential for the  generalisation of Denef's formalism with a fake field strength that is not a closed form.

As described in section \ref{sol-by-ans}, at least in principle these equations can be further simplified by eliminating fake charges $\Gamt$ from them and by using an ansatz $\cJ$ for the scalars, whereupon we obtain equations for $\cJ$. In particular, since we already have an ansatz \eqref{ansatz} for the stabilisation equations, we can determine $U$ and $\Omega$ from it, so that eq.~\eqref{eqmsimpleSimplified} becomes:
\begin{gather}
2\,\partial_{\tau}\I\bigl(\ee^{-U}\ee^{-\im\alpha}\Omega \bigr)=\partial_\tau\cJ\spc,\label{triv-flow}\\
-\Gamt = \partial_\tau\cJ-2\,\eta\,\R\bigl(\ee^{-U_\cJ}\ee^{-\im\alpha}\Omega_\cJ\bigr),\label{Gam-from-J}
\end{gather}
where $\eta$ is given by the static limit of \eqref{eta-ans} as
\begin{equation}\label{ans-def}
\eta=\ee^{2U_\cJ} \Iprod{\partial_\tau \cR}{\cH}\spc.
\end{equation}
Recall from the previous section that our ansatz is automatically a solution of the constraint \eqref{constr2} (and \eqref{etaS}), and the equations of motion are solved if one can find a $\cJ$, along the lines of section \ref{the-ansatz}, such that $\Gamt$ constructed above reproduces the black hole potential in \eqref{GamtGamt=V}, which now reads:
\begin{align}
\tfrac12 \lVert\partial_\tau\cJ\rVert^2 &= \tfrac12\lVert\Gamma\rVert^2 +\ee^{2U_\cJ}\Iprod{\cH}{\partial_\tau \cR}^2\spc.\label{eta-constr}
\end{align}
This quadratic constraint can be used to relate the physical charges to the harmonic functions $\cH$, in addition to the generic requirements of section \ref{the-ansatz}. In summary, the static equations of motion are integrable if there exists an $\cH$ and its corresponding $\cR$, constructed along the lines of section \ref{the-ansatz}, satisfying \eqref{eta-constr} .

This is similar in spirit, but different than the approach of \cite{Ceresole:2009iy, Bossard:2009we, Ceresole:2009vp}, were one seeks to rewrite the black hole potential in \eqref{GamtGamt=V} through a function of the physical charges and moduli $z^a$ directly. In contrast, \eqref{eta-constr} is an equation relating the harmonic functions controlling the physical charges to the ones controlling the scalars through the period vector $\Omega$.

We have checked that $\cJ$ for the known explicit static solutions are such that they satisfy \eqref{eta-constr} and hence are described by the flow equation \eqref{eqmsimpleSimplified} with $\eta$ as in \eqref{eta-ans}. Since all static non-BPS solutions are related by symplectic rotations to the seed solutions of section \ref{stab-ansatz}, it follows that they satisfy the same duality-covariant equations. The nontrivial $\eta$ is reflected in the anharmonic part of \eqref{stab-eqs-02}, controlled by the constant $b$ that remains after setting the angular momentum to zero in \eqref{har_fun_one}. This observation is in line with \cite{Gimon:2007mh}, where it was stressed that the crucial departure of the static non-BPS seed solution from a BPS-like ansatz is the presence of a parameter related to the asymptotic scalars, identified with this residual constant.

\subsection{The fake superpotential}\label{superpotential}

One can adopt the opposite point of view to the one taken above: first find fake charges $\Gamt$ reproducing the black hole potential \eqref{GamtGamt=V} and then solve the differential equations. Taking the intersection product of both sides of \eqref{eqmsimpleSimplified} with the basis elements then leads to the equations for the scale factor and the scalars, which, with appropriate identifications, have the form of non-supersymmetric flow equations generated by a superpotential $W$ \cite{Ceresole:2007wx}, analogous to the supersymmetric flow equations governed by the absolute value of the central charge:
\begin{align}
 \label{1_ord_UW}
\dot{U}=-\ee^U\ee^{-\im\alpha}\Iprod{\Gamt}{\Omega}&=-\ee^U W\spc,\\
 \label{1_order_zW}
\dot{z}^a=-\ee^U\ee^{-\im\alpha}g^{a\bar{b}} \Iprod{\Gamt}{\Dbar{b}{\Omega}}&=-2\ee^U g^{a\bar{b}}\bar{\partial}_{\bar{b}}W\spc.
\end{align}

Whenever $W$ is explicitly known for a given model and charge configuration, a practical way to connect it with our approach may be to first look for a moduli-independent matrix $\St$ that rotates the usual charge vector\footnote{By the shorthand $\St\Gamma$ we mean rotating the symplectic vector of charges corresponding to the coefficients of $\Gamma$: $S\cdot(p^I, q_J)^\mathrm{T}$, and arranging the result again as a three-form.} $\Gamma$ so that:
\begin{equation}\label{St}
 \lvert\tilde{Z}\rvert:=\lvert\Iprod{\St \Gamma}{\Omega}\rvert=W\spc.
\end{equation}
Then, its relation with $\Gamt$ is defined by:
\begin{equation}\label{Gamtil}
 \Gamt:= \im\bar{\tilde Z}\Omega-\im g^{\bar{a}b}\bar{\KD}_{\bar{a}}\bar{\tilde Z}\KD_b\Omega+\im g^{a\bar{b}}\KD_a \tilde Z\Dbar{b}{\Omega}-\im\tilde Z\bar{\Omega}\spc,
\end{equation}
where $\KD_a \tilde Z=\partial_a\tilde Z  +\frac12\partial_a K \tilde Z$, with $K$ being the K\"ahler potential. Note that in general $\Gamt\neq \St\Gamma$, if we allow for $\St$ to be complex, and in fact this turns out to be the simplest choice.

One can find the matrix $\St$ explicitly for the electric configuration, as in \eqref{stab-eqs-02}, assuming all physical scalars to have the same phase, say $f$. The relevant superpotential was given in \cite{Bellucci:2008sv}. Then, a suitable matrix $\St$ satisfying \eqref{St} and defining $\Gamt$ through \eqref{Gamtil} is
\begin{equation}\label{St-matrix}
\St = \operatorname{diag}\bigl(\ee^{-2\im f},1,1,1,\ee^{2\im f},1,1,1\bigr)\spc.
\end{equation}
In terms of the parameters appearing in the solution of section \ref{stab-ansatz} one can identify $\cot{f} = \ee^{2U} M$ and check that the equations of motion \eqref{eqmsimpleSimplified} are satisfied. The one-form $\boldeta$ is given by \eqref{eta-ans}.

In the non-supersymmetric axion-free case $M$ vanishes and $f=\pi/2$, so that $\St$ is constant (but not identity), while allowing for a $\tau$-dependent $f$ leads to more general non-supersymmetric solutions. It is worth noting that $\eta=0$ whenever $\St$ is constant (cf.~eq.~\eqref{etaS}). In particular, when $\St=\id$ we recover the supersymmetric case.

Alternatively, one can rewrite \eqref{Gam-from-J} and \eqref{eta-constr} in terms of a real matrix $\Sb$ such that:
\begin{align}
 \label{Gamma2}
\Sb\Gamma:=\Gamma_{\Sb} &= \tilde{\Gamma}-2\eta\R\bigl(\ee^{-U}\ee^{-\im\alpha}\Omega\bigr)\spc,\\
\tfrac12 \Iprod{\Gamma_\Sb}{\diamond\Gamma_\Sb} &=  V_\BH+\ee^{-2U}\eta^2\spc,\label{Fs_square}\\
\ee^{-\im\alpha}\Iprod{\Gamma_\Sb}{\Omega} &= W-\im\ee^{-U}\eta\spc,
\end{align}
where $W$ is the superpotential in \eqref{1_ord_UW}--\eqref{1_order_zW}. If $\Sb$ is known, it leads to simpler equations of motion for the scalars, that is
\begin{equation}\label{EOMShort}
 2\partial_{\tau}\I\bigl(\ee^{-U}\ee^{-\im\alpha}\Omega \bigr)=-\Gamma_\Sb\spc,
\end{equation}
which have the advantage of being directly integrable to \eqref{ansatz_rotBH}, giving the sought solutions as explained above. For the electric example above, $\Sb$ takes the form:
\begin{equation}\label{S-matr}
\Sb= \begin{pmatrix}
\mathbb{I} & 0 \\[.5ex]
\Ww & \mathbb{I}
\end{pmatrix}, \sep
\Ww= \begin{pmatrix}
-\frac{\ee^{-2U} \cot{f}}{(H^0)^2} & \frac{2q_a}{p^0} \\[2ex]
\frac{2q_a}{p^0} & 0
\end{pmatrix},
\end{equation}
so that \eqref{h-r-02} can be written as $\cJ=-\Sb\cH_c$, if the constants in \eqref{phys-charge} are appropriately chosen. Similarly to its complex counterpart $\St$, it reduces to a constant matrix in the axion-free case.

It is interesting to point out that the matrix \eqref{S-matr} is a (spacetime-dependent) element of the Peccei--Quinn group of transformations, defined as the largest subgroup of the symplectic group leaving the $X^I$'s and the K\"ahler potential invariant. As was shown recently \cite{Bellucci:2010zd}, applying such a transformation on the charges indeed shifts the black hole potential, as in \eqref{Fs_square}. For generic charges and phases of the scalars, the corresponding $\Sb$ can be found from the one in \eqref{S-matr} by conjugation with the appropriate element of the symplectic group. Such a matrix would leave a certain combination of $X^I$'s and $F_I$'s unchanged, e.g.~for the magnetic dual of the electric solution in \eqref{phys-charge} it would leave the $F_I$'s invariant. Identifying the combinations that must be invariant for a given set of charges could be a way to determine $\Sb$ from first principles.

\section{Conclusions and outlook}\label{Discussion}

In this work, we have extended the formalism of \cite{Denef:2000nb, Galli:2009bj}, deriving symplectically covariant flow equations for non-BPS extremal black holes in $\extN = 2$ supergravity, and we have constructed an ansatz for the corresponding stabilisation equations. The main novelty was to rewrite the electromagnetic part of the action in terms of a `fake' field strength two-form $\boldsymbol{\Ftilde}$ that does not have to be closed, where the non-closure turns out to be governed by a single one-form $\boldeta$. The presence of this one-form is further related to the axions in the full black hole solution, apparently rendering such a deformation essential in a general description of non-BPS extremal black holes. Unfortunately, this complication makes the full equations challenging to solve directly, at least without considerable intuition about the form of the solution.

To obtain that insight, we considered the known seed solution for under-rotating extremal black holes in theories with cubic prepotentials. We showed that it can indeed be written in terms of stabilisation equations for the period vector, just as BPS black holes. The crucial difference is that the scalars are not stabilised in terms of harmonic functions only, but one finds that a ratio of harmonic functions is required. When the angular momentum is set to zero, one simply has the inverse of a harmonic function, which vanishes near the horizon, but mixes with the other asymptotic constants at infinity.

As one might expect, a comparison of these explicit solutions with our flow equations reveals the non-closure of $\boldsymbol{\Ftilde}$ to be reflected in exactly these non-harmonic parts of the stabilisation equations. Based on this, we proposed an ansatz for the generic stabilisation equations of under-rotating extremal black holes, satisfying several requirements coming both from general arguments and known explicit solutions. Its practical realisation depends heavily on the model and in particular on invariants constructed from two charge vectors, one of which must correspond to a small black hole. Such invariants have been considered recently in \cite{Ferrara:2010cw, Ferrara:2010ug} in the context of multi-centre solutions.

In the static case, we showed how to combine this ansatz with the first-order flow equations to identify the structure of $\boldsymbol{\Ftilde}$ for known solutions and infer its general form. It turned out that there are two ways of connecting the result to previous work. One involves a complex matrix resembling the matrix $\St$ introduced in the superpotential formalism \cite{Ceresole:2007wx}. The other formulation simplifies the equations of motion, using the real matrix in \eqref{S-matr} belonging to the group of Peccei--Quinn transformations. This hints towards the possibility of obtaining such matrices systematically.

Irrespectively of the precise description, one can characterise any solution by a vector of harmonic functions such that the quartic invariant computed on their poles is related to the one associated to the physical charges by a sign flip. In principle, it is possible to construct this vector using purely algebraic methods, which is equivalent to solving the attractor equations for static non-supersymmetric black holes. Once this is known for particular charge vectors, one can replace the vector of charges with harmonic functions to find $\cH$ in \eqref{ansatz} and directly construct new solutions using our first-order equations.

We expect this to generally hold also for the single-centre rotating solutions covered by our ansatz, as the only difference with respect to static solutions is in the choice for the angular momentum harmonic function, without modifying the structure of \eqref{ansatz}. In line with this expectation, we have verified that the proposed ansatz does satisfy the constraint \eqref{Check}. However, it is more difficult to impose \eqref{constr1}, so it would be useful to find a generalisation of the arguments in section \ref{static-potential} and/or an extension of the fake superpotential formalism to the rotating case.

It is important to note that our flow equations are by construction fully covariant with respect to electric-magnetic duality and are compatible with the general seed solutions in four dimensions. It then follows that they capture the full orbit of non-BPS extremal solutions for suitable choices of $\boldsymbol{\Ftilde}$, regardless of the existence of other stationary points of the action, which should not be part of the standard non-BPS orbit of extremal black holes. It is also interesting to point out the similarity to special cases of static non-extremal black hole solutions, which can be obtained through a deformation of extremal solutions controlled by a ratio of harmonic functions \cite{Mohaupt:2010fk}, except that there it appears in the line element.

On the microscopic side, it would be very interesting to reproduce the stabilisation equations \eqref{ansatz}. In the rotating case the ratio of harmonic functions survives the near-horizon limit and modifies the attractor equations, similarly  to \cite{Astefanesei:2006dd}, so one generally expects this structure to be accessible from microscopics. Given the model of \cite{Gimon:2009gk}, where the constant part of $M$ in \eqref{h-r-02} is interpreted as the angle between wrapped D3 branes, one expects that the full angular momentum harmonic function might have a similar microscopic analogue.

Finally, it is worthwhile stressing that albeit the explicit solutions that we have discussed are only single centre, we have not made any assumptions on the number of centres in the derivation of flow equations in section \ref{eqns-from-action}. Also the ansatz \eqref{ansatz} is compatible with multi-centre harmonic functions. It would be illuminating to make a detailed comparison with the results of \cite{Bena:2009ev, Bena:2009en, Bena:2009fi}, as a test on the robustness of the assumption on the existence of stabilisation equations for generic extremal backgrounds.

\acknowledgments

We are very grateful to Dr A. Marrani for sharing with us a proof of the algebraic characterisation of doubly critical orbits mentioned in footnote \ref{orbit} and to an anonymous referee for his suggestions, which improved the presentation of the paper.

P.G. thanks Instituto de F\'{\i}sica Te\'{o}rica UAM/CSIC for hospitality and is grateful to M.A. Lled\'{o} for the help in the early stage of this project and the possibility to continue to work on it. The work of P.G. has been supported in part by grants FIS2008-06078-C03-02 and FPA2008-03811-E/INFN of Ministerio de Ciencia e Innovaci\'{o}n (Spain) and ACOMP/2010/213 from Generalitat Valenciana.

S.K. acknowledges fruitful discussions with B. de Wit throughout the course of this work and helpful comments on an earlier version of the manuscript. The work of S.K. is part of the research programme of the `Stichting voor Fundamenteel Onderzoek der Materie (FOM)', which is financially supported by the `Nederlandse Organisatie voor Wetenschappelijk Onderzoek (NWO)'. This work is also part of the ERC Advanced Grant research program no.\ 246974, \textit{``Supersymmetry: a window to non-perturbative physics''}.

J.P. wishes to thank Prof.~T.~Ort\'{\i}n for reading a draft of this paper and his valuable remarks, and Universiteit Utrecht for hospitality. J.P.'s work has been supported in part by the FWO-Vlaanderen project G.0235.05 and the Federal Office for Scientific, Technical and Cultural Affairs through the `Interuniversity Attraction Poles Programme--Belgian Science Policy' P6/11\nobreakdash-P, and in part by the Spanish Ministry of Science and Education grant FPA2009-07692, the Comunidad de Madrid grant HEPHACOS S2009ESP-1473 and the Spanish Consolider-Ingenio 2010 program CPAN CSD2007-00042.

\appendix

\section{On inverse harmonic functions}\label{inv-harm}

It is possible to give a generic heuristic argument for the presence of ratios of harmonic functions of the kind seen in \eqref{stab-eqs-02} in the solution for the scalars. We find it convenient to choose the arbitrary function $\ee^{\im\alpha}$ according to \eqref{Q+beta+eta+domega}, so that \eqref{Lmltc} reduces to
\begin{align}\label{stat-action}
\mathcal{L} =(\boldsymbol{\Gtilde}, \boldsymbol{\Gtilde})\spc.
\end{align}
Similarly to the gauge part of \eqref{S4Dmltc}, this can be interpreted as an action for the tensor
\begin{equation}
\tilde{\mathcal{G}} = \tilde{\mathcal{F}} - 2 \, \Im \boldsymbol{\Hodge_0} \mathbf{D} \Omt
                + 2 \,\Re \mathbf{D} (\ee^{2U} \Omt ) \wedge (\de t +\omega) \spc,\label{ps-tens}
\end{equation}
which respects the same pseudo-selfduality condition \eqref{W-def} as $\mathcal{F}$, assuming that $\tilde{\mathcal{F}}$ does:
\begin{equation}
\HM\Hodge\tilde{\mathcal{G}} = \tilde{\mathcal{G}}\spc.\label{ps-cons}
\end{equation}
The scalar part can be shown to be pseudo-selfdual using  \eqref{W-scal} and \eqref{CY-hodge}. The matrix $\HM$ is crucial for the existence of such a constraint, since it is not possible to impose selfduality on a four-dimensional field strength unless one complexifies it. However, it can be done in $4n+2$ dimensions, and \eqref{ps-cons} descends from the ten-dimensional constraint on the five-form.

For the ordinary Einstein--Maxwell theory, gauge field equations in backgrounds of the type \eqref{metricMltc} naturally lead to harmonic functions (cf.~eg.~appendix B in \cite{Mohaupt:2000mj}). Motivated by \eqref{scale-fact}, we further assume that the rotational one-form satisfies $\db \omega= \boldsymbol{\Hodge_0} \db M$. Denote spatial directions by $i=1,2,3$ and consider first electric solutions, $F_{ij}=0$, for which the Bianchi identity implies the existence of an electrostatic potential
\begin{equation}
\partial_i F_{tj} = \partial_{j} F_{ti} \sep \Rightarrow \sep F_{ti} = \partial_i \frac{M}{H}\spc.
\end{equation}
The field equations further impose that $H$ is harmonic
\begin{equation}
\ee^{-U} = H\spc, \sep \nabla^2 H = 0\spc.
\end{equation}
Similarly, for a magnetic solution, $F_{ti}=0$, the Bianchi identities relate the field strength to a harmonic function
\begin{equation}
\e^{tijk} \partial_{i} F_{jk}= 0\sep \Rightarrow \sep
F_{ij} =  \e_{ijk} \partial_k H \spc,
\end{equation}
where $\e_{ijk}$ is the Levi-Civita permutation symbol and the Einstein equation implies again $\ee^{-U} =H$. These solutions are related by an electric-magnetic duality rotation and belong to a class of solutions called the Majumdar--Papapetrou solutions.

Now, consider the case that the field strength is constrained to be pseudo-selfdual, as in \eqref{ps-cons}. Then, the distinction between equations of motion and Bianchi identities disappears, and the solutions can no longer be purely electric or purely magnetic. For such a field, the gauge part of the action in \eqref{S4Dmltc} leads to the equation of motion (see \cite{Bekaert:1998yp} for details)
\begin{equation}\label{self_du_gau}
\de(\mathcal{F} - \HM\Hodge\mathcal{F})=0\spc,
\end{equation}
which is then solved for \emph{both} cases:
\begin{align}
\boldsymbol{\mathcal{F}} = \boldsymbol{\Hodge_0}\db H\spc,\sep \ee^{-U} &= I(H)\spc,\label{self-sol}\\
\boldsymbol{\mathcal{F}} = \boldsymbol{\Hodge_0}\db \frac M H\spc,\sep \ee^{-U} &= I\Bigl(\frac M H\Bigr)\spc,\label{self-sol2}
\end{align}
where now $H$ denotes a vector of harmonic functions and $I$ is a model-dependent invariant. Imposing closure of $\boldsymbol{\mathcal F}$, which is equivalent to the existence of a gauge potential in four dimensions, one concludes that only the first solution survives, leading to the standard description by harmonic functions.

Going back to the Lagrangian in \eqref{stat-action}, the above discussion of selfdual fields applies including the scalar sector. In view of this, $\boldsymbol{\Gtilde}$, unlike $\sF$, is not necessarily closed and the two independent solutions in \eqref{self-sol}--\eqref{self-sol2} are allowed. Based on this, one concludes that a vector of inverse harmonic functions is a zero mode of the equations of motion following from \eqref{stat-action}. Such a vector must be part of the general solution for the scalar sector only, given that the gauge field $\mathcal{F}$ is described by harmonic functions, as above.

Since the equations of motion are nonlinear, it is a nontrivial task to find full solutions with both kinds of zero modes turned on. Nevertheless, this discussion demonstrates that one can consider the scalar sector a priori independently from the vector one. It also lends credibility to the presence of ratios of harmonic functions in the ansatz \eqref{ansatz} and the different sets of harmonic functions in \eqref{stab-eqs-02}.

\bibliographystyle{utphys}
\bibliography{References}

\providecommand{\href}[2]{#2}\begingroup\raggedright\begin{thebibliography}{10}

\bibitem{Ferrara:1995ih}
S.~Ferrara, R.~Kallosh, and A.~Strominger, ``{$N=2$} extremal black holes,''
  \href{http://dx.doi.org/10.1103/PhysRevD.52.R5412}{{\em Phys. Rev.}
  {\bfseries D52} (1995) 5412--5416},
\href{http://arxiv.org/abs/hep-th/9508072}{{\ttfamily arXiv:hep-th/9508072}}.

\bibitem{Strominger:1996kf}
A.~Strominger, ``Macroscopic entropy of {$N=2$} extremal black holes,''
  \href{http://dx.doi.org/10.1016/0370-2693(96)00711-3}{{\em Phys. Lett.}
  {\bfseries B383} (1996) 39--43},
\href{http://arxiv.org/abs/hep-th/9602111}{{\ttfamily arXiv:hep-th/9602111}}.

\bibitem{Ferrara:1996dd}
S.~Ferrara and R.~Kallosh, ``Supersymmetry and attractors,''
  \href{http://dx.doi.org/10.1103/PhysRevD.54.1514}{{\em Phys. Rev.} {\bfseries
  D54} (1996) 1514--1524},
\href{http://arxiv.org/abs/hep-th/9602136}{{\ttfamily arXiv:hep-th/9602136}}.

\bibitem{Behrndt:1997ny}
K.~Behrndt, D.~L{\"u}st, and W.~A. Sabra, ``Stationary solutions of {$N=2$}
  supergravity,'' \href{http://dx.doi.org/10.1016/S0550-3213(97)00633-0}{{\em
  Nucl. Phys.} {\bfseries B510} (1998) 264--288},
\href{http://arxiv.org/abs/hep-th/9705169}{{\ttfamily arXiv:hep-th/9705169}}.

\bibitem{Denef:2000nb}
F.~Denef, ``Supergravity flows and {D}-brane stability,''
  \href{http://dx.doi.org/10.1088/1126-6708/2000/08/050}{{\em JHEP} {\bfseries
  08} (2000) 050},
\href{http://arxiv.org/abs/hep-th/0005049}{{\ttfamily arXiv:hep-th/0005049}}.

\bibitem{Cardoso:2000qm}
G.~L. Cardoso, B.~de~Wit, J.~K{\"a}ppeli, and T.~Mohaupt, ``Stationary {BPS}
  solutions in {$N=2$} supergravity with {$R^2$} interactions,''
  \href{http://dx.doi.org/10.1088/1126-6708/2000/12/019}{{\em JHEP} {\bfseries
  12} (2000) 019},
\href{http://arxiv.org/abs/hep-th/0009234}{{\ttfamily arXiv:hep-th/0009234}}.

\bibitem{Gauntlett:2002nw}
J.~P. Gauntlett, J.~B. Gutowski, C.~M. Hull, S.~Pakis, and H.~S. Reall, ``All
  supersymmetric solutions of minimal supergravity in five dimensions,''
  \href{http://dx.doi.org/10.1088/0264-9381/20/21/005}{{\em Class. Quant.
  Grav.} {\bfseries 20} (2003) 4587--4634},
\href{http://arxiv.org/abs/hep-th/0209114}{{\ttfamily arXiv:hep-th/0209114}}.

\bibitem{Gauntlett:2004qy}
J.~P. Gauntlett and J.~B. Gutowski, ``General concentric black rings,''
  \href{http://dx.doi.org/10.1103/PhysRevD.71.045002}{{\em Phys. Rev.}
  {\bfseries D71} (2005) 045002},
\href{http://arxiv.org/abs/hep-th/0408122}{{\ttfamily arXiv:hep-th/0408122}}.

\bibitem{Castro:2008ne}
A.~Castro, J.~L. Davis, P.~Kraus, and F.~Larsen, ``String theory effects on
  five-dimensional black hole physics,''
  \href{http://dx.doi.org/10.1142/S0217751X08039724}{{\em Int. J. Mod. Phys.}
  {\bfseries A23} (2008) 613--691},
\href{http://arxiv.org/abs/0801.1863}{{\ttfamily arXiv:0801.1863 [hep-th]}}.

\bibitem{Khuri:1995xq}
R.~R. Khuri and T.~Ort{\'\i}n, ``A non-supersymmetric dyonic extreme
  {R}eissner--{N}ordstr{\"o}m black hole,''
  \href{http://dx.doi.org/10.1016/0370-2693(96)00139-6}{{\em Phys. Lett.}
  {\bfseries B373} (1996) 56--60},
\href{http://arxiv.org/abs/hep-th/9512178}{{\ttfamily arXiv:hep-th/9512178}}.

\bibitem{Ferrara:1997tw}
S.~Ferrara, G.~W. Gibbons, and R.~Kallosh, ``Black holes and critical points in
  moduli space,'' \href{http://dx.doi.org/10.1016/S0550-3213(97)00324-6}{{\em
  Nucl. Phys.} {\bfseries B500} (1997) 75--93},
\href{http://arxiv.org/abs/hep-th/9702103}{{\ttfamily arXiv:hep-th/9702103}}.

\bibitem{Sen:2005wa}
A.~Sen, ``Black hole entropy function and the attractor mechanism in higher
  derivative gravity,''
  \href{http://dx.doi.org/10.1088/1126-6708/2005/09/038}{{\em JHEP} {\bfseries
  09} (2005) 038},
\href{http://arxiv.org/abs/hep-th/0506177}{{\ttfamily arXiv:hep-th/0506177}}.

\bibitem{Goldstein:2005hq}
K.~Goldstein, N.~Iizuka, R.~P. Jena, and S.~P. Trivedi, ``Non-supersymmetric
  attractors,'' \href{http://dx.doi.org/10.1103/PhysRevD.72.124021}{{\em Phys.
  Rev.} {\bfseries D72} (2005) 124021},
\href{http://arxiv.org/abs/hep-th/0507096}{{\ttfamily arXiv:hep-th/0507096}}.

\bibitem{Tripathy:2005qp}
P.~K. Tripathy and S.~P. Trivedi, ``Non-supersymmetric attractors in string
  theory,'' \href{http://dx.doi.org/10.1088/1126-6708/2006/03/022}{{\em JHEP}
  {\bfseries 03} (2006) 022},
\href{http://arxiv.org/abs/hep-th/0511117}{{\ttfamily arXiv:hep-th/0511117}}.

\bibitem{Kallosh:2006ib}
R.~Kallosh, N.~Sivanandam, and M.~Soroush, ``Exact attractive non-{BPS} {STU}
  black holes,'' \href{http://dx.doi.org/10.1103/PhysRevD.74.065008}{{\em Phys.
  Rev.} {\bfseries D74} (2006) 065008},
\href{http://arxiv.org/abs/hep-th/0606263}{{\ttfamily arXiv:hep-th/0606263}}.

\bibitem{Cardoso:2007ky}
G.~L. Cardoso, A.~Ceresole, G.~Dall'Agata, J.~M. Oberreuter, and J.~Perz,
  ``First-order flow equations for extremal black holes in very special
  geometry,'' \href{http://dx.doi.org/10.1088/1126-6708/2007/10/063}{{\em JHEP}
  {\bfseries 10} (2007) 063},
\href{http://arxiv.org/abs/0706.3373}{{\ttfamily arXiv:0706.3373 [hep-th]}}.

\bibitem{Gimon:2007mh}
E.~G. Gimon, F.~Larsen, and J.~Sim{\'o}n, ``Black holes in supergravity: the
  non-{BPS} branch,''
  \href{http://dx.doi.org/10.1088/1126-6708/2008/01/040}{{\em JHEP} {\bfseries
  01} (2008) 040},
\href{http://arxiv.org/abs/0710.4967}{{\ttfamily arXiv:0710.4967 [hep-th]}}.

\bibitem{Bellucci:2008sv}
S.~Bellucci, S.~Ferrara, A.~Marrani, and A.~Yeranyan, ``{$stu$} black holes
  unveiled,'' \href{http://dx.doi.org/10.3390/e10040507}{{\em Entropy}
  {\bfseries 10} (2008) 507--555},
\href{http://arxiv.org/abs/0807.3503}{{\ttfamily arXiv:0807.3503 [hep-th]}}.

\bibitem{Ferrara:1997uz}
S.~Ferrara and M.~G{\"u}naydin, ``Orbits of exceptional groups, duality and
  {BPS} states in string theory,''
  \href{http://dx.doi.org/10.1142/S0217751X98000913}{{\em Int. J. Mod. Phys.}
  {\bfseries A13} (1998) 2075--2088},
\href{http://arxiv.org/abs/hep-th/9708025}{{\ttfamily arXiv:hep-th/9708025}}.

\bibitem{Bellucci:2006xz}
S.~Bellucci, S.~Ferrara, M.~G{\"u}naydin, and A.~Marrani, ``Charge orbits of
  symmetric special geometries and attractors,''
  \href{http://dx.doi.org/10.1142/S0217751X06034355}{{\em Int. J. Mod. Phys.}
  {\bfseries A21} (2006) 5043--5098},
\href{http://arxiv.org/abs/hep-th/0606209}{{\ttfamily arXiv:hep-th/0606209}}.

\bibitem{Ferrara:2007tu}
S.~Ferrara and A.~Marrani, ``On the moduli space of non-{BPS} attractors for
  {$N=2$} symmetric manifolds,''
  \href{http://dx.doi.org/10.1016/j.physletb.2007.07.001}{{\em Phys. Lett.}
  {\bfseries B652} (2007) 111--117},
\href{http://arxiv.org/abs/0706.1667}{{\ttfamily arXiv:0706.1667 [hep-th]}}.

\bibitem{Astefanesei:2006dd}
D.~Astefanesei, K.~Goldstein, R.~P. Jena, A.~Sen, and S.~P. Trivedi, ``Rotating
  attractors,'' \href{http://dx.doi.org/10.1088/1126-6708/2006/10/058}{{\em
  JHEP} {\bfseries 10} (2006) 058},
\href{http://arxiv.org/abs/hep-th/0606244}{{\ttfamily arXiv:hep-th/0606244}}.

\bibitem{Rasheed:1995zv}
D.~Rasheed, ``The rotating dyonic black holes of {K}aluza--{K}lein theory,''
  \href{http://dx.doi.org/10.1016/0550-3213(95)00396-A}{{\em Nucl. Phys.}
  {\bfseries B454} (1995) 379--401},
\href{http://arxiv.org/abs/hep-th/9505038}{{\ttfamily arXiv:hep-th/9505038}}.

\bibitem{Matos:1996km}
T.~Matos and C.~Mora, ``Stationary dilatons with arbitrary electromagnetic
  field,'' \href{http://dx.doi.org/10.1088/0264-9381/14/8/027}{{\em Class.
  Quant. Grav.} {\bfseries 14} (1997) 2331--2340},
\href{http://arxiv.org/abs/hep-th/9610013}{{\ttfamily arXiv:hep-th/9610013}}.

\bibitem{Larsen:1999pp}
F.~Larsen, ``Rotating {K}aluza--{K}lein black holes,''
  \href{http://dx.doi.org/10.1016/S0550-3213(00)00064-X}{{\em Nucl. Phys.}
  {\bfseries B575} (2000) 211--230},
\href{http://arxiv.org/abs/hep-th/9909102}{{\ttfamily arXiv:hep-th/9909102}}.

\bibitem{Goldstein:2008fq}
K.~Goldstein and S.~Katmadas, ``Almost {BPS} black holes,''
  \href{http://dx.doi.org/10.1088/1126-6708/2009/05/058}{{\em JHEP} {\bfseries
  05} (2009) 058},
\href{http://arxiv.org/abs/0812.4183}{{\ttfamily arXiv:0812.4183 [hep-th]}}.

\bibitem{Bena:2009ev}
I.~Bena, G.~Dall'Agata, S.~Giusto, C.~Ruef, and N.~P. Warner, ``Non-{BPS} black
  rings and black holes in {T}aub-{NUT},''
  \href{http://dx.doi.org/10.1088/1126-6708/2009/06/015}{{\em JHEP} {\bfseries
  06} (2009) 015},
\href{http://arxiv.org/abs/0902.4526}{{\ttfamily arXiv:0902.4526 [hep-th]}}.

\bibitem{Bena:2009en}
I.~Bena, S.~Giusto, C.~Ruef, and N.~P. Warner, ``Multi-center non-{BPS} black
  holes---the solution,''
  \href{http://dx.doi.org/10.1088/1126-6708/2009/11/032}{{\em JHEP} {\bfseries
  11} (2009) 032},
\href{http://arxiv.org/abs/0908.2121}{{\ttfamily arXiv:0908.2121 [hep-th]}}.

\bibitem{Bena:2009qv}
I.~Bena, S.~Giusto, C.~Ruef, and N.~P. Warner, ``A (running) bolt for new
  reasons,'' \href{http://dx.doi.org/10.1088/1126-6708/2009/11/089}{{\em JHEP}
  {\bfseries 11} (2009) 089},
\href{http://arxiv.org/abs/0909.2559}{{\ttfamily arXiv:0909.2559 [hep-th]}}.

\bibitem{Bena:2009fi}
I.~Bena, S.~Giusto, C.~Ruef, and N.~P. Warner, ``Supergravity solutions from
  floating branes,'' \href{http://dx.doi.org/10.1007/JHEP03(2010)047}{{\em
  JHEP} {\bfseries 03} (2010) 047},
\href{http://arxiv.org/abs/0910.1860}{{\ttfamily arXiv:0910.1860 [hep-th]}}.

\bibitem{Breitenlohner:1987dg}
P.~Breitenlohner, D.~Maison, and G.~W. Gibbons, ``Four-dimensional black holes
  from {K}aluza--{K}lein theories,''
\href{http://dx.doi.org/10.1007/BF01217967}{{\em Commun. Math. Phys.}
  {\bfseries 120} (1988) 295}.

\bibitem{Gunaydin:2005mx}
M.~G{\"u}naydin, A.~Neitzke, B.~Pioline, and A.~Waldron, ``{BPS} black holes,
  quantum attractor flows and automorphic forms,''
  \href{http://dx.doi.org/10.1103/PhysRevD.73.084019}{{\em Phys. Rev.}
  {\bfseries D73} (2006) 084019},
\href{http://arxiv.org/abs/hep-th/0512296}{{\ttfamily arXiv:hep-th/0512296}}.

\bibitem{Gaiotto:2007ag}
D.~Gaiotto, W.~Li, and M.~Padi, ``Non-supersymmetric attractor flow in
  symmetric spaces,''
  \href{http://dx.doi.org/10.1088/1126-6708/2007/12/093}{{\em JHEP} {\bfseries
  12} (2007) 093},
\href{http://arxiv.org/abs/0710.1638}{{\ttfamily arXiv:0710.1638 [hep-th]}}.

\bibitem{Bergshoeff:2008be}
E.~Bergshoeff, W.~Chemissany, A.~Ploegh, M.~Trigiante, and T.~Van~Riet,
  ``Generating geodesic flows and supergravity solutions,''
  \href{http://dx.doi.org/10.1016/j.nuclphysb.2008.10.023}{{\em Nucl. Phys.}
  {\bfseries B812} (2009) 343--401},
\href{http://arxiv.org/abs/0806.2310}{{\ttfamily arXiv:0806.2310 [hep-th]}}.

\bibitem{Bossard:2009at}
G.~Bossard, H.~Nicolai, and K.~S. Stelle, ``Universal {BPS} structure of
  stationary supergravity solutions,''
  \href{http://dx.doi.org/10.1088/1126-6708/2009/07/003}{{\em JHEP} {\bfseries
  07} (2009) 003},
\href{http://arxiv.org/abs/0902.4438}{{\ttfamily arXiv:0902.4438 [hep-th]}}.

\bibitem{Bossard:2009my}
G.~Bossard and H.~Nicolai, ``Multi-black holes from nilpotent {L}ie algebra
  orbits,'' \href{http://dx.doi.org/10.1007/s10714-009-0870-2}{{\em Gen. Rel.
  Grav.} {\bfseries 42} (2010) 509--537},
\href{http://arxiv.org/abs/0906.1987}{{\ttfamily arXiv:0906.1987 [hep-th]}}.

\bibitem{Bossard:2010mv}
G.~Bossard, ``$1/8$ {BPS} black hole composites,''
\href{http://arxiv.org/abs/1001.3157}{{\ttfamily arXiv:1001.3157 [hep-th]}}.

\bibitem{Ceresole:2007wx}
A.~Ceresole and G.~Dall'Agata, ``Flow equations for non-{BPS} extremal black
  holes,'' \href{http://dx.doi.org/10.1088/1126-6708/2007/03/110}{{\em JHEP}
  {\bfseries 03} (2007) 110},
\href{http://arxiv.org/abs/hep-th/0702088}{{\ttfamily arXiv:hep-th/0702088}}.

\bibitem{Andrianopoli:2007gt}
L.~Andrianopoli, R.~D'Auria, E.~Orazi, and M.~Trigiante, ``First order
  description of black holes in moduli space,''
  \href{http://dx.doi.org/10.1088/1126-6708/2007/11/032}{{\em JHEP} {\bfseries
  11} (2007) 032},
\href{http://arxiv.org/abs/0706.0712}{{\ttfamily arXiv:0706.0712 [hep-th]}}.

\bibitem{Perz:2008kh}
J.~Perz, P.~Smyth, T.~Van~Riet, and B.~Vercnocke, ``First-order flow equations
  for extremal and non-extremal black holes,''
  \href{http://dx.doi.org/10.1088/1126-6708/2009/03/150}{{\em JHEP} {\bfseries
  03} (2009) 150},
\href{http://arxiv.org/abs/0810.1528}{{\ttfamily arXiv:0810.1528 [hep-th]}}.

\bibitem{Ceresole:2009iy}
A.~Ceresole, G.~Dall'Agata, S.~Ferrara, and A.~Yeranyan, ``First order flows
  for {$N=2$} extremal black holes and duality invariants,''
  \href{http://dx.doi.org/10.1016/j.nuclphysb.2009.09.003}{{\em Nucl. Phys.}
  {\bfseries B824} (2010) 239--253},
\href{http://arxiv.org/abs/0908.1110}{{\ttfamily arXiv:0908.1110 [hep-th]}}.

\bibitem{Bossard:2009we}
G.~Bossard, Y.~Michel, and B.~Pioline, ``Extremal black holes, nilpotent orbits
  and the true fake superpotential,''
  \href{http://dx.doi.org/10.1007/JHEP01(2010)038}{{\em JHEP} {\bfseries 01}
  (2010) 038},
\href{http://arxiv.org/abs/0908.1742}{{\ttfamily arXiv:0908.1742 [hep-th]}}.

\bibitem{Ceresole:2009vp}
A.~Ceresole, G.~Dall'Agata, S.~Ferrara, and A.~Yeranyan, ``Universality of the
  superpotential for {$d = 4$} extremal black holes,''
  \href{http://dx.doi.org/10.1016/j.nuclphysb.2010.02.015}{{\em Nucl. Phys.}
  {\bfseries B832} (2010) 358--381},
\href{http://arxiv.org/abs/0910.2697}{{\ttfamily arXiv:0910.2697 [hep-th]}}.

\bibitem{Ceresole:2010hq}
A.~Ceresole and S.~Ferrara, ``Black holes and attractors in supergravity,''
\href{http://arxiv.org/abs/1009.4175}{{\ttfamily arXiv:1009.4175 [hep-th]}}.

\bibitem{Miller:2006ay}
C.~M. Miller, K.~Schalm, and E.~J. Weinberg, ``Nonextremal black holes are
  {BPS},'' \href{http://dx.doi.org/10.1103/PhysRevD.76.044001}{{\em Phys. Rev.}
  {\bfseries D76} (2007) 044001},
\href{http://arxiv.org/abs/hep-th/0612308}{{\ttfamily arXiv:hep-th/0612308}}.

\bibitem{Galli:2009bj}
P.~Galli and J.~Perz, ``Non-supersymmetric extremal multicenter black holes
  with superpotentials,'' \href{http://dx.doi.org/10.1007/JHEP02(2010)102}{{\em
  JHEP} {\bfseries 02} (2010) 102},
\href{http://arxiv.org/abs/0909.5185}{{\ttfamily arXiv:0909.5185 [hep-th]}}.

\bibitem{Andrianopoli:2009je}
L.~Andrianopoli, R.~D'Auria, E.~Orazi, and M.~Trigiante, ``First order
  description of {$D=4$} static black holes and the {H}amilton--{J}acobi
  equation,'' \href{http://dx.doi.org/10.1016/j.nuclphysb.2010.02.020}{{\em
  Nucl. Phys.} {\bfseries B833} (2010) 1--16},
\href{http://arxiv.org/abs/0905.3938}{{\ttfamily arXiv:0905.3938 [hep-th]}}.

\bibitem{Chemissany:2010zp}
W.~Chemissany, P.~Fr{\'e}, J.~Rosseel, A.~S. Sorin, M.~Trigiante, and
  T.~Van~Riet, ``Black holes in supergravity and integrability,''
  \href{http://dx.doi.org/10.1007/JHEP09(2010)080}{{\em JHEP} {\bfseries 09}
  (2010) 080},
\href{http://arxiv.org/abs/1007.3209}{{\ttfamily arXiv:1007.3209 [hep-th]}}.

\bibitem{deWit:1984pk}
B.~de~Wit and A.~Van~Proeyen, ``Potentials and symmetries of general gauged
  {$N=2$} supergravity: {Y}ang--{M}ills models,''
\href{http://dx.doi.org/10.1016/0550-3213(84)90425-5}{{\em Nucl. Phys.}
  {\bfseries B245} (1984) 89}.

\bibitem{deWit:1984px}
B.~de~Wit, P.~G. Lauwers, and A.~Van~Proeyen, ``Lagrangians of {$N=2$}
  supergravity-matter systems,''
\href{http://dx.doi.org/10.1016/0550-3213(85)90154-3}{{\em Nucl. Phys.}
  {\bfseries B255} (1985) 569}.

\bibitem{Ceresole:1995ca}
A.~Ceresole, R.~D'Auria, and S.~Ferrara, ``The symplectic structure of {$N=2$}
  supergravity and its central extension,''
  \href{http://dx.doi.org/10.1016/0920-5632(96)00008-4}{{\em Nucl. Phys. Proc.
  Suppl.} {\bfseries 46} (1996) 67--74},
\href{http://arxiv.org/abs/hep-th/9509160}{{\ttfamily arXiv:hep-th/9509160}}.

\bibitem{Henneaux:1988gg}
M.~Henneaux and C.~Teitelboim, ``Dynamics of chiral (self-dual) {$p$}-forms,''
\href{http://dx.doi.org/10.1016/0370-2693(88)90712-5}{{\em Phys. Lett.}
  {\bfseries B206} (1988) 650}.

\bibitem{Bekaert:1998yp}
X.~Bekaert and M.~Henneaux, ``Comments on chiral {$p$}-forms,''
  \href{http://dx.doi.org/10.1023/A:1026610530708}{{\em Int. J. Theor. Phys.}
  {\bfseries 38} (1999) 1161--1172},
\href{http://arxiv.org/abs/hep-th/9806062}{{\ttfamily arXiv:hep-th/9806062}}.

\bibitem{Nampuri:2010um}
S.~Nampuri and M.~Soroush, ``New perspectives on attractor flows and trees from
  {CFT},''
\href{http://arxiv.org/abs/1009.5768}{{\ttfamily arXiv:1009.5768 [hep-th]}}.

\bibitem{Shmakova:1996nz}
M.~Shmakova, ``{C}alabi--{Y}au black holes,''
  \href{http://dx.doi.org/10.1103/PhysRevD.56.540}{{\em Phys. Rev.} {\bfseries
  D56} (1997) 540--544},
\href{http://arxiv.org/abs/hep-th/9612076}{{\ttfamily arXiv:hep-th/9612076}}.

\bibitem{Behrndt:2005he}
K.~Behrndt, G.~L. Cardoso, and S.~Mahapatra, ``Exploring the relation between
  {4D} and {5D} {BPS} solutions,''
  \href{http://dx.doi.org/10.1016/j.nuclphysb.2005.10.026}{{\em Nucl. Phys.}
  {\bfseries B732} (2006) 200--223},
\href{http://arxiv.org/abs/hep-th/0506251}{{\ttfamily arXiv:hep-th/0506251}}.

\bibitem{Gaiotto:2005gf}
D.~Gaiotto, A.~Strominger, and X.~Yin, ``New connections between {4D} and {5D}
  black holes,'' \href{http://dx.doi.org/10.1088/1126-6708/2006/02/024}{{\em
  JHEP} {\bfseries 02} (2006) 024},
\href{http://arxiv.org/abs/hep-th/0503217}{{\ttfamily arXiv:hep-th/0503217}}.

\bibitem{Gaiotto:2005xt}
D.~Gaiotto, A.~Strominger, and X.~Yin, ``{5D} black rings and {4D} black
  holes,'' \href{http://dx.doi.org/10.1088/1126-6708/2006/02/023}{{\em JHEP}
  {\bfseries 02} (2006) 023},
\href{http://arxiv.org/abs/hep-th/0504126}{{\ttfamily arXiv:hep-th/0504126}}.

\bibitem{Cerchiai:2009pi}
B.~L. Cerchiai, S.~Ferrara, A.~Marrani, and B.~Zumino, ``Duality, entropy and
  {ADM} mass in supergravity,''
  \href{http://dx.doi.org/10.1103/PhysRevD.79.125010}{{\em Phys. Rev.}
  {\bfseries D79} (2009) 125010},
\href{http://arxiv.org/abs/0902.3973}{{\ttfamily arXiv:0902.3973 [hep-th]}}.

\bibitem{Nampuri:2007gv}
S.~Nampuri, P.~K. Tripathy, and S.~P. Trivedi, ``On the stability of
  non-supersymmetric attractors in string theory,''
  \href{http://dx.doi.org/10.1088/1126-6708/2007/08/054}{{\em JHEP} {\bfseries
  08} (2007) 054},
\href{http://arxiv.org/abs/0705.4554}{{\ttfamily arXiv:0705.4554 [hep-th]}}.

\bibitem{Mohaupt:2000mj}
T.~Mohaupt, ``Black hole entropy, special geometry and strings,'' {\em Fortsch.
  Phys.} {\bfseries 49} (2001) 3--161,
\href{http://arxiv.org/abs/hep-th/0007195}{{\ttfamily arXiv:hep-th/0007195}}.

\bibitem{Ceresole:2010nm}
A.~Ceresole, S.~Ferrara, and A.~Marrani, ``Small {$N=2$} extremal black holes
  in special geometry,''
  \href{http://dx.doi.org/10.1016/j.physletb.2010.08.053}{{\em Phys. Lett.}
  {\bfseries B693} (2010) 366--372},
\href{http://arxiv.org/abs/1006.2007}{{\ttfamily arXiv:1006.2007 [hep-th]}}.

\bibitem{Ferrara:2010ug}
S.~Ferrara, A.~Marrani, E.~Orazi, R.~Stora, and A.~Yeranyan, ``Two-center black
  holes duality-invariants for {$stu$} model and its lower-rank descendants,''
\href{http://arxiv.org/abs/1011.5864}{{\ttfamily arXiv:1011.5864 [hep-th]}}.

\bibitem{Marrani:priv}
A.~Marrani (private communication).

\bibitem{Bates:2003vx}
B.~Bates and F.~Denef, ``Exact solutions for supersymmetric stationary black
  hole composites,''
\href{http://arxiv.org/abs/hep-th/0304094}{{\ttfamily arXiv:hep-th/0304094}}.

\bibitem{Bellucci:2010zd}
S.~Bellucci, A.~Marrani, and R.~Roychowdhury, ``Topics in cubic special
  geometry,''
\href{http://arxiv.org/abs/1011.0705}{{\ttfamily arXiv:1011.0705 [hep-th]}}.

\bibitem{Ferrara:2010cw}
S.~Ferrara, A.~Marrani, and E.~Orazi, ``Split attractor flow in {$N=2$}
  minimally coupled supergravity,''
  \href{http://dx.doi.org/10.1016/j.nuclphysb.2011.01.015}{{\em Nucl. Phys.}
  {\bfseries B846} (2011) 512--541},
\href{http://arxiv.org/abs/1010.2280}{{\ttfamily arXiv:1010.2280 [hep-th]}}.

\bibitem{Mohaupt:2010fk}
T.~Mohaupt and O.~Vaughan, ``Non-extremal black holes, harmonic functions, and
  attractor equations,''
  \href{http://dx.doi.org/10.1088/0264-9381/27/23/235008}{{\em Class. Quant.
  Grav.} {\bfseries 27} (2010) 235008},
\href{http://arxiv.org/abs/1006.3439}{{\ttfamily arXiv:1006.3439 [hep-th]}}.

\bibitem{Gimon:2009gk}
E.~G. Gimon, F.~Larsen, and J.~Sim{\'o}n, ``Constituent model of extremal
  non-{BPS} black holes,''
  \href{http://dx.doi.org/10.1088/1126-6708/2009/07/052}{{\em JHEP} {\bfseries
  07} (2009) 052},
\href{http://arxiv.org/abs/0903.0719}{{\ttfamily arXiv:0903.0719 [hep-th]}}.

\end{thebibliography}\endgroup

\end{document}